\documentclass[fleqn,usenatbib]{mnras}

\usepackage[T1]{fontenc}
\usepackage{setspace}
\DeclareRobustCommand{\VAN}[3]{#2}
\let\VANthebibliography\thebibliography
\def\thebibliography{\DeclareRobustCommand{\VAN}[3]{##3}\VANthebibliography}

\usepackage[flushleft]{threeparttable}

\usepackage{graphicx}	% Including figure files
\usepackage{amsmath}	% Advanced maths commands
\usepackage{amssymb}	% Extra maths symbols
\usepackage{longtable}
\usepackage{float}
\newcommand{\vplp}{$\nu_\mathrm{peak}-L_\mathrm{peak} $}

\usepackage{newtxtext,newtxmath}
\title[]{The Relativistic Jet Dichotomy and the End of the Blazar Sequence}

\author[M. Keenan et al.]{
Mary Keenan,$^{1}$
Eileen T. Meyer,$^{1}$\thanks{meyer@umbc.edu}
Markos Georganopoulos,$^{1,2}$
Karthik Reddy$^{1}$,
Omar J. French$^{1}$
\\
$^{1}$Department of Physics, University of Maryland Baltimore County, Baltimore, MD 21250, USA\\
$^{2}$NASA Goddard Space Flight Center, Code 663, Greenbelt, MD 20771, USA\\
}

\date{Accepted 2021 April 19. Received 2021 April 16; in original form 27 July 2020}

\pubyear{2021}

\begin{document}
\label{firstpage}
\pagerange{\pageref{firstpage}--\pageref{lastpage}}
\maketitle

\begin{abstract}
Our understanding of the unification of jetted AGN has evolved greatly as jet samples have increased in size.
%(e.g. 132 sources for the blazar sequence of \citealt{fossati98}, 257 for the blazar envelope of \citealt{meyer2011}).  
Here, based on the largest-ever sample of over 2000 well-sampled jet spectral energy distributions, we examine the synchrotron peak frequency -- peak luminosity plane, and find little evidence for the anti-correlation known as the blazar sequence. Instead, we find strong evidence for a dichotomy in jets, between those associated with efficient or `quasar-mode' accretion (strong/type II jets) and those associated with inefficient accretion (weak/type I jets). Type II jets include those hosted by high-excitation radio galaxies, flat-spectrum radio quasars (FSRQ), and most low-frequency-peaked BL Lac objects. Type I jets include those hosted by low-excitation radio galaxies and blazars with synchrotron peak frequency above $10^{15}$\,Hz (nearly all BL Lac objects). We have derived estimates of the total jet power for over 1000 of our sources from low-frequency radio observations, and find that the jet dichotomy does \emph{not} correspond to a division in jet power. Rather, type II jets are produced at \emph{all} observed jet powers, down to the lowest levels in our sample, while type I jets range from very low to moderately high jet powers, with a clear upper bound at $\sim10^{43}$\,erg\,s$^{-1}$. The range of jet power in each class matches exactly what is expected for efficient (i.e., a few to 100\% Eddington) or inefficient ($<0.5$\% Eddington) accretion onto black holes ranging in mass from $10^7-10^{9.5}\,M_\odot$. % The neat separation of strong and weak blazars in the synchrotron peak frequency -- peak luminosity plane is not yet fully understood, but is likely related to the presence or absence of external photon fields as well as the difference between their velocity structures.

\end{abstract}


%%%%%%%%%%%%%%%%%%%%%%%%%%%%%%%%%%%%%%%%%%%%%%%%%%

%%%%%%%%%%%%%%%%% BODY OF PAPER %%%%%%%%%%%%%%%%%%

\section{Introduction}
\label{sec:intro}

Radio-Loud Active Galactic Nuclei (RL AGN) exhibit highly collimated relativistic jets of non-thermal plasma originating very near the central super-massive black hole (of $10^6-10^{10} M_\odot$) and propagating out to kpc - Mpc scales \citep[see, e.g.][for a recent review]{blandford2019}. They can have a major impact on their host galaxy and surrounding environment, heating the intercluster medium \citep{mcnamara07,2012ApJ...752...23C} and halting or (more rarely) initiating star formation \citep{cattaneo09,2019ApJ...881..147S}.  For several decades it has been known that strong radio emission from a small fraction of AGN signals the presence of a relativistic jet \citep[e.g.][]{1995ApJ...438...62W}, but it is still unknown why only some AGN produce them. While all AGN host an actively accreting super-massive black hole \citep{1991Natur.349..138R}, there are only a few properties which describe it: mass, spin, accretion rate and mode, and spin orientation. Each of these (along with environment) has been posited as a possible determiner of being jetted versus non-jetted, with no clear consensus \citep[e.g.][]{2010ApJ...711...50T,2010MNRAS.406..975G,chiaberge2011,2012MNRAS.419L..69N,chiaberge2015,2020ApJ...890..144G}. Within the subset of jetted AGN, these same properties are also considered possible determinants of the morphological divide in radio jets first noted by Fanaroff \& Riley in the 1970s \cite[e.g.][]{ghisellini01,2003ApJ...593..667M,2004MNRAS.351..733M,2010MNRAS.406..975G,2014Natur.515..376G}.

%The study of jetted AGN is complicated by the extreme variety of appearance of these objects, both in the sense of how they appear in resolved radio images and in their optical and broadband spectral properties. 
Historically, a large number of sub-classes of radio-loud AGN have been defined, usually based on observational properties in the band in which they were discovered -- e.g. steep-spectrum radio quasars, optically violent variable sources, X-ray selected BL~Lacertae objects, broad-line radio galaxies, and many more \citep[see ][]{urry95}. Part of the extreme variety of appearance is clearly due to differences in viewing angle, as the radiation from the jet comes to dominate the spectral energy distribution (SED) when oriented at small angles due to Doppler boosting, which can enhance the apparent luminosity of the source by factors of hundreds to thousands, obscuring the host galaxy and other spectral components \citep{1979ApJ...232...34B,markos98,2012MNRAS.420.2899G}. Putting aside the myriad and overlapping historical classes, \emph{observationally} the divide to first order is between sources classified as radio galaxies (sources with jets pointed $>10-15$ degrees away) and blazars, where the jet is aligned to within a few degrees to the line-of-sight \citep{1990ApJ...356...75P,1992ApJ...387..449P}. The exact dividing line between the two types is not very well-defined since it is difficult to estimate the exact viewing angle of a source, and classifications are often somewhat arbitrary in the transition zone \citep[e.g.][]{3c264_veritas_2020}.

A long-standing open question is what drives the wide range of phenomenology in jetted AGN that is unrelated to viewing angle. Based on some of the earliest high-resolution radio surveys, \cite{fanaroff74} observed a clear division in radio morphology and power, between lower-power `plume-like' jets that fade with distance (FR type I), and powerful and highly collimated jets ending in bright hotspots (FR II).  While FR I jets are on average less powerful than FR II jets, there is substantial overlap, and deeper surveys have revealed that low-power jets can be either FR I or II \citep{2017A&A...601A..81C,2019MNRAS.488.2701M}. It is possible that the FR I/II morphological divide is related to the way that these jets are formed, i.e., through either the  Blandford-Znajek \cite[BZ,][]{blandford77} process which taps the spin energy of the black hole, or the Blandford-Payne \cite[BP,][]{1982MNRAS.199..883B} process which extracts rotational energy through the accretion disk. Recent observations of the black hole shadow in M87, which hosts an FR I jet, as an example, appear consistent with the BZ process\citep[e.g.][]{2019ApJ...875L...6E}.
%The more powerful FR II jets are typically hosted in systems with signs of an efficient accretion mode, such as a quasi-thermal "big blue bump" (BBB) in the UV and strong emission lines \citep{2004ApJ...614...91W}.
%[I don't think this is a good citation here] This has been confirmed by recent VLBI observations of Cygnus A \citep[FR II,][]{2017Galax...5...22B}. 
%In contrast, observations by the Event Horizon Telescope of the black hole shadow in M87, a well-known FR I, are consisent with a jet powered by the BZ process
% \citep{2019ApJ...875L...6E}, and FR I jets generally do not show signs of efficient or highly accreting black holes. Careful observations by \cite{croston18} have shown that the slowed plasma filling the space around FR I jets contains a high proton fraction, suggesting entrained material, while FR II jets only contain electrons, indicating that they did not entrain material from the surrounding medium and were not disrupted.  This assumes that the jet starts with only leptons, in agreement with the BZ process. 
The difference could also be in part environmental; while FR I and FR II jets have been found to reside in similar types of galaxies \citep[i.e. luminous early-type galaxies with large black hole masses, ][]{2017A&A...598A..49C,2017A&A...601A..81C}, FR I jets are more often found in denser environments, such as clusters \citep{2011AJ....141...88W,2019A&A...622A..10C}. 

Jetted AGN, as in the larger class of AGN generally, can also be divided based on their optical spectra into high-excitation or low-excitation radio galaxies (HERG or LERG, respectively) based on the prominence of several of their spectral lines \citep{best2012}. The division is thought to be the result of differences in accretion mode, with HERGs occurring in systems with radiatively efficient and therefore more luminous accretion disks \citep[e.g.][]{shakura73} while LERGs are the result of radiatively inefficient disks \citep[e.g.][]{narayan95}. More powerful radio sources are thought to have more luminous accretion disks, which are able to activate more line emission from the BLR, resulting in a higher excitation state. The correspondence of power and spectral type in early studies of blazars lead to the suggestion that the low-power and line-less blazars known as BL Lac objects (BL Lacs) might correspond to FR I radio galaxies \citep{1990ApJ...356...75P} while high-power and broad-lined blazars (now known generally as flat-spectrum radio quasars or FSRQ) might correspond to FR II sources \citep{1992ApJ...387..449P}.  The early small samples of FR I and FR II radio galaxies also had a strong overlap with LERGs and HERGs respectively \citep{laing94,best2012}. More recent studies with larger and deeper surveys shows much less correspondence between the FR morphological classes and spectral types \citep{2017MNRAS.466.4346M}. Nonetheless, an idea of a "zeroth order" unification scheme where BL Lacs and FR Is are LERGs and FSRQ and FR II are HERGs, has generally prevailed.

In blazars, the SED exhibits a characteristic `double-peaked' spectrum, where the first broad peak (occurring between $10^{12}$ to $10^{18}$ Hz) is due to Doppler-boosted synchrotron emission, and the second (peaking between $10^{18}$ and $10^{27}$\,Hz) is due to inverse Compton emission, where either synchrotron or external photons are upscattered by the relativistic particles in the jet \citep[e.g.][]{sikora94}. 
Somewhat in contrast to the idea of a dichotomy of jetted AGN, even early studies of blazars found a broad continuum in source characteristics, such as synchrotron peak frequency and broad-band spectral indices \citep{giommi1994,sambruna1996}. In what has become one of the most cited early works on blazar populations, \cite{fossati98} discovered an anti-correlation between the synchrotron peak luminosity ($L_\textrm{peak}$)\footnote{or equivalently, the 5 GHz radio luminosity} and the synchrotron peak frequency ($\nu_\textrm{peak}$). Based on a sample of 48 X-ray selected and 84 radio-selected blazars, this observed anti-correlation has become known as the blazar sequence (by analogy to the stellar main sequence). The blazar sequence put in order the significant range in synchrotron peak frequencies (which range from sub-mm to hard X-rays) and implied that jets are essentially monoparametric, with the key parameter being the radio luminosity (or equivalently, the total jet power). While the blazar sequence result has been challenged by observations of outliers and evidence of selection effects \citep[e.g.][]{2002babs.conf..133G,2003ApJ...588..128P,2006A&A...445..441N,padovani2012,2012A&A...541A.160G,2016A&A...587A...8R,cerruti2017}, it is still frequently adopted as a description of blazars as a population \citep[e.g.][]{2012Sci...338.1190A}.  %

%Whether there is an \emph{intrinsic} dichotomy in the class of jetted AGN, or simply a continuous sequence in some parameter space, is the subject of this paper.
%[Problems with unification can go here, I would not mention hybrid morphology sources here -- put that in the discussion perhaps. A better thing to focus on would be the lack of a clear transition from FR I to II at a certain power,  the problem of fake BL Lacs, the fact that FR I have bigger black holes (I think), the herg/lerg mismatch, and the need for velocity gradients (chiaberge 2001)] 2003ApJ...588..128P,2012MNRAS.422L..48P,cerruti2017

In studies of blazars it is frequently invoked  \cite[including in ][and elsewhere]{fossati98} that most blazars are observed at small angles with respect to the line of sight, and that the viewing angle is of minor importance in the phenomenology.
However, this is not the case as noted by e.g., \cite{markos98} and \cite{2008MNRAS.387.1669G}.
The need to consider orientation angle as well as the importance of using an estimate of jet power not confused by relativistic beaming was the motivation for the re-examination of the blazar sequence in \citealt{meyer2011}, hereafter M11. In that work, the following scenario was tested: could the total jet power be the primary parameter governing jet properties, as mass is for the stellar main sequence? It was also expected that with deeper surveys, a `blazar envelope' would develop beneath the original blazar sequence (itself formed from the most aligned sources) due to the presence of more misaligned jets with lesser Doppler boosting. The main finding in M11 was that while more misaligned jets did fill in an envelope below the blazar sequence, there were also signs of a dichotomy, leading to the classification of `strong' and `weak' jets.\footnote{The equivalent classes are renamed `type II' and `type I' in this paper.} The strong jets included powerful  BL Lacs with `low' synchrotron peak frequencies ($<10^{15}$~Hz) and practically all FSRQ and FR IIs, while the weak jets included  BL Lacs at higher peak frequencies ($>10^{15}$~Hz) and FR I radio galaxies. There was also some suggestion that `sequence-like' behavior, where peak frequency was driven by jet power, might exist within the strong and weak classes separately \citep{2011arXiv1111.4711G,2012arXiv1205.0794M}.

Nearly 10 years have passed since the work of M11 and the number of surveys and amount of available data has increased significantly. A number of publications have also noted the existence of `high-frequency, high-luminosity' peaked blazars that appear inconsistent with the blazar sequence/envelope \citep{2003ApJ...588..128P,2012MNRAS.422L..48P,cerruti2017}. In this paper, our aim is to re-examine blazar phenomenology and in particular the evidence for a dichotomy in the jet population and its link to accretion. The general approach we have taken in this study is similar to M11, in that our aim is to collect the largest possible sample of well-characterized jet SEDs, in order to get the most complete picture of the phenomenology of the population.  We have opted for this approach because it is better suited to revealing the characteristics of the full population, and because looking at the largest possible sample is the best way to answer questions of which parts of the parameter space are filled or empty. While there are advantages to using only statistically complete samples, there are also major disadvantages -- namely, that such samples are small, and almost none have well-constrained broad-band SEDs for the entire sample as we require here.  In the next section we describe the extensive data collection efforts that have lead to the characterization of thousands of jet SEDs. In Section~\ref{sec:results} we present our new results and discuss the implications, and finally summarize our conclusions in Section~\ref{sec:conclusions}. 

In this paper we have adopted a standard $\Lambda$-cosmology, with $\Omega_{M} = 0.308$, $\Omega_{\Lambda} = 0.692$, and $H_{0} = 67.8$ km s\textsuperscript{-1} Mpc\textsuperscript{-1}\citep{planck16}. Spectral indices are defined such that $f_{\nu} \propto \nu^{-\alpha}$ unless otherwise noted.  Quantities written as luminosities ($L$) are references to powers ($\nu L_{\nu}$ implied).

\begin{table*}
    \centering
    \caption {Radio-Loud AGN Samples} 
    \begin{tabular}{l|c|c|c|c|c|c|c}
        Sample & Abbr. & $N_\mathrm{init}$ & $N_\mathrm{final}$ & $N_\mathrm{unique}$ & Reference & Ref. Let.\\
        (1)&(2)&(3)&(4) & (5) & (6) & (7)\\
        \hline                            
                            1 Jansky Blazar Sample &       1Jy &   34 &   34 &    0 &                                  \cite{stickel91} & a \\
          2 Jansky Survey of Flat-Spectrum Sources &       2Jy &  232 &  129 &    5 &                       \cite{1985MNRAS.216..173W}  & b \\
                   2-degree field (2dF) QSO survey &       2QZ &   26 &    1 &    0 &   \cite{2002MNRAS.334..941L,2007MNRAS.374..556L}  & c \\
        Third Cambridge Catalogue of Radio Sources &      3CRR &  173 &   68 &    2 &                       \cite{1983MNRAS.204..151L}  & d \\
         The 3rd Catalog of Hard Fermi-LAT Sources &      3FHL & 1553 &  446 &    1 &                      \cite{2017ApJS..232...18A}   & e \\
  The 3rd Catalog of AGN Detected by the Fermi/LAT &      3LAC & 1894 &  838 &  100 &                                   \cite{3lac}     & f \\
                Molonglo Equatorial Radio Galaxies &       ... &  178 &   49 &    8 &                                    \cite{best99}  & g \\
             The Candidate Gamma-Ray Blazar Survey &    CGRaBs & 1625 & 1154 &  467 &                       \cite{2008ApJS..175...97H}  & h \\
                        Cosmic Lens All-Sky Survey &     CLASS &  232 &   79 &   23 &                       \cite{2004yCat..73480937C}  & i \\
                    Deep X-Ray Radio Blazar Survey &     DXRBS &  283 &  151 &   47 &                           \cite{landt2001_dxrbs}  & j \\
     Einstein Slew Survey Sample of BL Lac Objects &       ... &   66 &   14 &    0 &                       \cite{1996ApJS..104..251P}  & k \\
              Hamburg-RASS Bright X-ray AGN Sample &       HRX &  172 &   42 &    2 &                        \cite{2003AA...401..927B}  & l \\
          The MOJAVE Sample of VLBI Monitored Jets &       ... &  512 &  456 &   21 &                            \cite{lister2018}$^2$  & m \\
    Metsahovi Radio Observatory BL Lacertae sample &       ... &  393 &  168 &    3 &                        \cite{2006AA...445..441N}  & n \\
        Parkes Quarter-Jansky Flat-Spectrum Sample &       ... &  878 &  524 &  155 &                        \cite{2002AA...386...97J}  & o \\
                Radio-Emitting X-ray Source Survey &       REX &  143 &   32 &    9 &                       \cite{1999ApJ...513...51C}  & p \\
                       Radio-Optical-X-ray Catalog &      ROXA &  801 &  234 &  101 &                        \cite{2007AA...472..699T}  & q \\
                   RASS - Green Bank BL Lac sample &       RGB &  127 &   72 &    1 &                       \cite{1999ApJ...525..127L}  & r \\
     Einstein Medium-Sensitivity Survey of BL Lacs &      EMSS &   52 &    6 &    0 &                       \cite{2000AJ....120.1626R}  & s \\
                  RASS - SDSS Flat-Spectrum Sample &       ... &  501 &   96 &   21 &                       \cite{2008AJ....135.2453P}  & t \\
             Sedentary Survey of High-Peak BL Lacs &       ... &  150 &   27 &    1 &                       \cite{1999MNRAS.310..465G}  & u \\
                Ultra Steep Spectrum Radio Sources &       ... &  668 &    2 &    0 &                        \cite{2000AAS..143..303D}  & v \\
                         The X-Jet Online Database &       ... &  117 &   96 &   10 &                                            $^2$   & w \\
        \hline
$^1$http://www.physics.purdue.edu/astro/MOJAVE/allsources.html \\
$^2$https://hea-www.harvard.edu/XJET/ \\
    \end{tabular}
    \label{blazar_samples}
\end{table*}

\section{Methods}
\label{sec:methods}
 We first describe our initial sample of jetted sources in Section~\ref{sec:init}, followed by the extensive data collection efforts we have undertaken to build multi-wavelength SEDs in Section~\ref{sec:data}. In Section~\ref{sec:lowfreq} we describe our method of estimating the jet power from radio observations, and in Section~\ref{sec:sedfit} we describe our method of broad-band SED fitting, leading to the final samples (Sections~\ref{sec:final}$-$\ref{sec:speeds}).

\subsection{Initial Sample Selection}
\label{sec:init}
The initial sample of jetted AGN was compiled from the catalogs of jetted sources listed in Table~\ref{blazar_samples}, where we give the catalog name, common abbreviation, total number of sources in that sample ($N_\mathrm{init}$), the total from that sample included in our `well-sampled' (TEX/UEX) catalog ($N_\mathrm{final}$) as well as the number of unique sources contributed to the latter ($N_\mathrm{unique}$), and the sample reference. Column 7 gives a unique letter code which is used in the later catalog tables to identify which samples an individual source appears in. The total sample comprises 6856 sources after accounting for duplicates.  Many sources in this list have very poorly sampled SEDs, often with little to no data beyond the radio. We have attempted to gather as much archival photometric and/or imaging data as possible at all wavelengths for this initial sample, and have also conducted observing campaigns in the radio and sub-mm in order to maximize the subset of the sample with well-sampled SEDs. 

\subsection{Data Collection}
\label{sec:data}
%We describe here the published archives and catalogs we have utilized in this effort, as well as the new and archival data which was specifically analyzed for our sample. 
In addition to the published catalogs and newly reduced data described below, we also utilized observations from our Sub-milliter Array (SMA) projects 2018A-S027 and 2018B-S003, as well as data from the VLA Low-band Ionosphere and Transient Experiment (VLITE) which is managed by the Naval Research Lab\footnote{https://www.nrl.navy.mil/rsd/7210/7213/vlite}.  
%A description of the data reduction for these observations is described in Keenan et al. (2021), in preparation. 

\subsubsection{Catalogs and Published Observations}
\label{sec:catalog}
The main source of published photometry measurements was the NASA Extragalactic Database\footnote{https://ned.ipac.caltech.edu/} (NED).  As the intent is to fit the SED of the non-thermal jet or radio lobes, all data from NED was automatically filtered to exclude emission line and host galaxy fluxes and the SED for each source was inspected by eye prior to fitting for outliers.  We also take flux measurements from various surveys and other publications not included in NED (30 total). These sources are listed in Table \ref{imported_data}, with the sample name in column 1, the abbreviation of the sample in column 2 and reference in column 3. The source names in each catalog were matched to our initial sample using the SIMBAD\footnote{http://simbad.u-strasbg.fr/simbad/sim-fid} astronomical database and CDS XMatch.\footnote{http://cdsxmatch.u-strasbg.fr/}

\begin{table*}
\caption {Published Data Sources} 
\label{imported_data}
\begin{tabular}{l c c}
%\toprule
Name & Abbr. & Reference\\ 
(1)&(2)&(3)\\
\hline
AllWISE Catalog of Mid-IR AGNs  &...  & \cite{allwiseagn} \\
ALMA Calibrator Continuum Observations Catalog & ALMACAL    & \cite{almacal}    \\
The Chandra Source Catalog &... & \cite{chandra}    \\
The First Brazil-ICRANet Gamma-ray Blazar Catalogue & 1BIGB &  \cite{2018MNRAS.480.2165A} \\
Fermi Large Area Telescope Third Source Catalog & 3FGL  & \cite{3fgl}       \\
The GAIA Mission &GAIA & \cite{gaia}       \\
Galaxy And Mass Assembly &GAMA  & \cite{gama}       \\
The Galactic and Extra-Galactic All-sky MWA Survey  & GLEAM & \cite{gleamcat}   \\
The GMRT 150 MHz all-sky radio survey & TGSS & \cite{gmrttgss}   \\
The Herschel Astrophysical Terahertz Large Area Survey & H-ATLAS  & \cite{hatlas}     \\
The LOFAR Two-metre Sky Survey & LoTSS   & \cite{lotss}      \\
Planck 2013 results  &...& \cite{planck}     \\ 
The Spitzer Mid-Infrared Active Galactic Nucleus Survey &...& \cite{spitzer} \\
SWIFT BAT 105-month survey &... & \cite{bat105}     \\
SWIFT X-ray telescope point source catalogue & 1SXPS   & \cite{xrt1}       \\
The Third Catalog of Hard Fermi-LAT Sources & 3FHL  & \cite{2017ApJS..232...18A}       \\
The Very Large Array Low-frequency Sky Survey Redux & VLSSr & \cite{vlssr}      \\
Published Radio Maps  &...&\begin{tabular}{@{}c@{}}\cite{kharb10,ant85}; \\ \cite{landt08,murphy93}\\\cite{cassaro99}\end{tabular}\\
Published Nuclear Fluxes & ... & \begin{tabular}{@{}c@{}}\cite{chiaberge2002,2000AA...362..871C}; \\ \cite{2003AA...403..889T,2004AA...428..401V}\\\cite{chiaberge1999,2003MNRAS.338..176H}\\\cite{2018ApJS..234....7M,2015ApJS..220....5M}\end{tabular}\\
\hline
%\bottomrule
\end{tabular}
\end{table*}

%\subsubsection{New and Previously Unpublished Archival Observations}

%[ETM note: you need to describe in numbers what NEW data was analyzed. You have new GMRT, VLA, SMA, and my ALMA programs that need to be somewhat described! see vla section. Also make sure there are the proper NRAO acknowledgements and data source statements at the end of the paper.]

%** what about SMA?? **

\subsubsection{Very Large Array}
We have analyzed 434 archival observations from the pre-upgrade Very Large Array (VLA) and 26 observations from the upgraded Karl G. Jansky Very Large Array (JVLA) which were not previously published. 
%We also analyze 5 new L-band JVLA observations taken for this project under program 19A-288.  
Observations were reduced using the Common Astronomy Software Applications (CASA) package \citep{mcmullin2007}.
Details of these observations can be found in Table \ref{vla-table}, which lists the source name in column 1, band in column 2, frequency of observation in column 3, project code in column 4 and observation date in column 5.   
%If the CASA VLA calibration pipeline was used, the version number is listed in column 6, otherwise it is listed as `n/a'.  
The RMS of the final image in Jy/beam is listed in column 6, beam size in column 7, and core flux in Jy in column 8.  

For the historical VLA data, we followed a standard data reduction in CASA. For most observations, 3C~286, 3C~48, or 3C~147 served as the flux calibrator and the source itself as the phase calibrator.  For JVLA observations, we used the CASA pipeline to obtain the initial calibrated measurement set.  Images were produced using the CLEAN deconvolution algorithm as implemented in CASA, with Briggs weighting and robust parameter of 0.5. For JVLA data, we used \texttt{nterms=2} in the deconvolution to account for spectral curvature over the wide observing band.  In general, the calibration was improved through several rounds of (non-cumulative) phase-only self-calibration followed by a single round of amplitude and phase self-calibration, with rare exceptions where the source was not strong enough to self-calibrate. In the final images, the sources generally appeared as point sources, or else there was a clear point-source `core' easily visible against background extended emission. The core flux was measured using the Gaussian fit feature in CASA.

\begin{table*}
\caption {VLA Observations} 
\label{vla-table}
\begin{tabular}{l c c c c c  c c}
%\toprule
\centering
Source & Band & Frequency & Project Code & Observation Date & RMS & Beam Size & Core Flux \\
--- & --- & GHz & --- & YYYY-MM-DD & Jy/beam & arcsec & Jy 
\\ 
(1)&(2)&(3)&(4)&(5)&(6)&(7)&(8)\\%\midrule
\hline
             1Jy 0748+126   &       C   &    4.86   &       AM0672   &   2000-11-07   &   1.35e-04   &    0.54 x  0.40   &   1.36e+00\\
             1Jy 0748+126   &       L   &    1.43   &       AM0672   &   2000-11-05   &   8.36e-04   &    1.63 x  1.32   &   1.39e+00\\
                   3C 189   &       L   &    1.43   &       AL0604   &   2004-10-25   &   1.04e-04   &    1.44 x  1.24   &   1.32e-01\\
              PKS 0754+10   &       L   &    1.56   &       AB0310   &   1984-12-24   &   7.25e-03   &    1.53 x  1.47   &   1.70e+00\\
                   3C 191   &       C   &    4.75   &       AK0180   &   1987-07-26   &   6.40e-05   &    0.47 x  0.41   &   5.70e-02\\
\hline \\
%bottomrule
\end{tabular}
\begin{tablenotes}
\item Table \ref{vla-table} is published in its entirety in the machine-readable format.  A portion is shown here for guidance regarding its form and content.
\end{tablenotes}
\end{table*}

\subsubsection{Atacama Large Millimeter/submillimeter Array}
We analyzed 116 archival observations from the Atacama Large Millimeter/submillimeter Array (ALMA) for this project.  We also analyzed 16, 25, and 12  new observations from our projects 2015.1.00932.S, 2016.1.01481.S, and 2017.1.01572.S, respectively. The details of the observations are given in Table \ref{alma-table}, which lists the source name in column 1, band in column 2, frequency of observation in column 3, project code in column 4, observation date in column 5, the RMS of the final image in Jy/beam in column 6, beam size in arcseconds in column 7, and core flux in Jy in column 8.  To produce the initial calibrated measurement set, we used the CASA pipeline versions listed in column 6 to run the provided \texttt{scriptForPI.py}, which gives a science-ready measurement set. We then produced images using the CLEAN algorithm with Briggs weighting, robust parameter of 0.5 and \texttt{nterms=2}. Imaging was improved by applying several rounds of phase-only self-calibration and a single round of amplitude and phase self-calibration to the measurement set.  We then measured the core flux and beam size using the Gaussian fit feature in CASA.

%142 total ALMA observations

\begin{table*}
\caption {ALMA Observations} 
\label{alma-table}
\begin{tabular}{l c c c c c c c}
%\toprule
\centering
Source & Band & Frequency & Project Code & Observation Date & RMS & Beam Size & Core Flux \\
--- & --- & GHz & --- & YYYY-MM-DD & Jy/beam & arcsec & Jy 
\\
(1)&(2)&(3)&(4)&(5)&(6)&(7)&(8)\\%\midrule
\hline
              1Jy 0742+10   &       3   &    95.99   &   2016.1.01567.S   &   2016-11-06   &   2.47e-04   &    0.70 x 0.63   &   2.52e-01\\
                   3C 189   &       6   &   221.00   &   2015.1.00324.S   &   2016-06-18   &   1.96e-04   &    1.04 x 0.49   &   6.62e-02\\
              1Jy 0834-20   &       3   &    89.06   &   2015.1.00856.S   &   2016-03-15   &   1.13e-04   &    2.11 x 1.60   &   6.25e-01\\
                 4C 29.30   &       6   &   233.00   &   2017.1.01572.S   &   2017-12-31   &   4.48e-04   &    6.80 x 6.05   &   4.00e-03\\
                   3C 212   &       3   &    97.50   &   2019.1.01709.S   &   2019-10-14   &   4.37e-05   &    1.18 x 1.07   &   7.52e-02\\
\hline \\
%bottomrule
\end{tabular}
\begin{tablenotes}
\item Table \ref{alma-table} is published in its entirety in the machine-readable format.  A portion is shown here for guidance regarding its form and content.
\end{tablenotes}
\end{table*}

\subsubsection{Hubble Space Telescope Observations}
\label{hst_fluxes}
%This project requires a well-sampled jet (core) spectrum in order to fit the nonthermal SED; however, t
The SED of radio galaxies is usually dominated by non-jet sources beyond the radio -- e.g., the host galaxy, infrared dust emission, and potentially the accretion disk. In order to accurately fit the non-thermal jet spectrum in radio galaxies, we used published core jet fluxes isolated in high-resolution imaging by The Hubble Space Telescope and  Chandra X-ray Observatory \citep{chiaberge2002,2000AA...362..871C,2003AA...403..889T,2004AA...428..401V,chiaberge1999,2003MNRAS.338..176H,2018ApJS..234....7M,2015ApJS..220....5M}. In addition we utilized archival HST observations available in the Hubble Legacy Archive (HLA) for (otherwise well-sampled) radio galaxies without published nuclear/core optical fluxes. The HLA provides science-ready images, and a total of 8 observations were analyzed using the \texttt{astropy} module in Python.  For each image, the host galaxy was fit with a 2D S\'ersic model after masking the core region (selected by hand).  The best-fit S\'ersic model was then subtracted from the unmasked image and the core emission was measured, where we assume that the excess flux in the central core region is attributed to the jet.  The resulting fluxes are given in Table \ref{hst_table}, which lists the source name in column 1, observation date in column 2, instrument and filter in column 3, frequency of observation in column 4 and nuclear flux in column 5.

\begin{table}
    \centering
    \caption{HST Archival Data}
    \begin{tabular}{l|c|c|c|r}
        Name & Date & Inst./Filter & Freq.y & Flux  \\
        -- & -- & -- & Hz & Jy \\
        (1)&(2)&(3)&(4)&(5)\\
        \hline
                           3C 208 &     1994-04-04 &  WFPC2/F702W & 4.28e14 & 1.31e-04 \\
                   3C 287 &     1994-03-05 &  WFPC2/F702W &     4.28e14 & 2.44e-04 \\
                   3C 196 &     1994-04-16 &  WFPC2/F702W &     4.28e14 & 1.83e-04 \\
                   3C 336 &     1994-09-12 &  WFPC2/F702W &     4.28e14 & 1.35e-04 \\
                   3C 438 &     1994-12-15 &  WFPC2/F702W &     4.28e14 & $<$ 2.13e-05 \\
                   3C 192 &     1997-01-13 &  WFPC2/F555W &     5.40e14 & 1.49e-04 \\
                    3C 41 &     1994-07-29 &  WFPC2/F555W &     5.40e14 & 3.01e-06 \\
                 3C 427.1 &     1994-03-23 &  WFPC2/F702W &     4.28e14 & 1.03e-06 \\
        \hline \\
    \end{tabular}
    \label{hst_table}
\end{table}

\subsection{Jet Power}
\label{sec:lowfreq}

\begin{figure}
    \centering
   \includegraphics[width=0.45\textwidth]{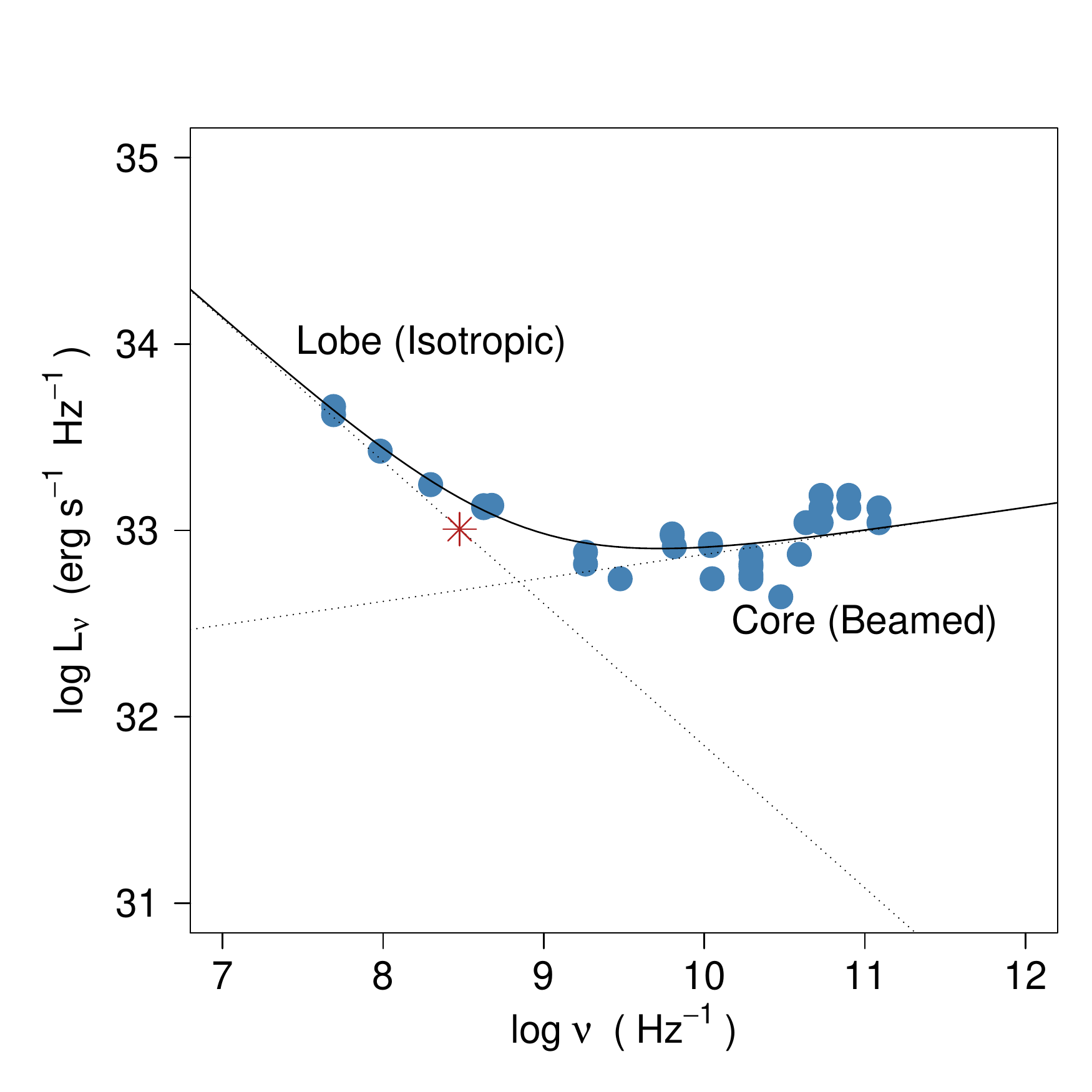}
    \caption{The low-frequency SED of blazar TXS 1700+685 consists of a steep component due to isotropically emitting lobe emission, and the flat component due to the point source core.  Here the spectrum is clearly best-fit with an added (double) power-law. The dotted black lines show the two components, while the solid black line shows the total fit.  All flux measurements are shown as blue circles. The red star shows our estimate of $L_\textrm{ext}$ taken at 300 MHz.}
    \label{sd_example}
\end{figure}

In most jetted AGN, the low-frequency SED generally consists of two components: a steep component due to the extended (isotropic) emission which dominates at low frequencies and a flat component due to the beamed point source core of the jet which dominates at high frequencies (Figure \ref{sd_example}). The steep component is due to the isotropically emitting slowed plasma in the radio lobes.  In this study, $L_\textrm{ext}$ is the observed luminosity of the steep component measured at 300 MHz. It has been shown that $L_\textrm{ext}$ correlates with estimates of the lifetime-integrated jet power \citep[e.g.][]{cavagnolo10,godfrey13,ineson17,birzan20}. In comparison, the radio core luminosity is not a good indicator of the true jet power, as it is enhanced by orders of magnitude from the intrinsic value due to Doppler boosting, and in general there is a strong degeneracy between angle, speed, and the intrinsic luminosity. The extended radio emission is not a perfect measure of jet power, as there is considerable dispersion around the scaling relations referenced above, and it is almost certainly not reliable for individual sources \citep{2013MNRAS.430..174H}. 
%$L_\textrm{ext}$ represents an average of the jet power over timescales of a few tens to hundreds of Myr (and may even be the average over multiple episodes of jet activity). In comparison, the core luminosity is an instantaneous representation of the jet power. So for a restarted jet where the lobe has faded away, the core would be extremely luminous but without any lobe emission. Similarly, for a turned off jet, the lobe may be extremely luminous but without any emission from the core. It has been statistically proven that these two quantities are in fact tail-independent \citep{2018ApJS..239...33Y}.  
However, it is the only estimate of jet power which we can easily obtain for a large sample (over a thousand sources as noted below), and with appropriate caution it should be suitable for broad characterization of large samples.
%although in combination with other parameters, it can be used to estimate the orientation of the source, as discussed in \ref{orientation_indicator}.  
In this analysis we have assumed that contributions from extended jets or hotspots to $L_\textrm{ext}$ are minimal; this has been verified through careful checking of several dozen sources with these features.  

One method to measure $L_\textrm{ext}$ involves decomposing the spectrum by looking for a break in the spectral index, as depicted in Figure \ref{sd_example}.  The individual components are well-described by power laws over the frequency ranges in which they dominate, and the total spectrum can be fit with an added power law.  For each source, we initially took all data below $10^{13}$ Hz and compared single and added power-law fits to the data. As these are nested models, with the added power law having two additional degrees of freedom (the normalization and index of the second power law), a $\chi^2$ test with 2 degrees of freedom was used.  We accepted the added power law for a significance level $\alpha > 0.95$, indicating that both components were reliably detected and measured.  If rejected, then we conclude that only one component (well-fit with a single power law) is present.  In cases where the core spectrum begins to show curvature at high frequencies, we reduced the maximum frequency (from $10^{13}$ Hz) to an appropriate value so that the core could be well-described by a power law.  

The spectral decomposition method is most easily utilized in moderately misaligned sources.  In aligned sources, the core emission can dominate down to the very lowest radio frequencies observed.  In these cases the two components can still be separately fit using measurements from deep, high-resolution radio maps, where the extended emission can be imaged and measured directly. Several prior works (listed in Table \ref{imported_data}) have used this method; we also derived extended flux measurements from our own radio maps (187 observations). The 300-MHz flux can then be estimated either through a fit to two or more extended flux measurements or by extrapolating from a single point to 300 MHz with a spectral index of 0.7 (e.g., M11). In some (rare) cases, only an extended component is detected by the combined fitting (meaning the source is very misaligned and the core de-boosted so that the extended emission dominates the total flux), and high-resolution radio maps were used to measure the core flux directly.

Once the core and lobe spectral fits were determined, we used them to obtain the quantities of interest: the extended flux at 300 MHz ($L_\textrm{ext}$), the (beamed) core flux at 1.4 GHz ($L_{core}$), and the radio core dominance ($R_\textrm{c}$), which we defined as the log of the ratio of the core to lobe luminosity at 1.4 GHz.  We also computed the crossing frequency, $\nu_\textrm{cross}$, which is the frequency at which the spectrum transitions from extended-dominated to core-dominated (i.e. where the fits intersect, as depicted in Figure \ref{sd_example}). 

In sources where the extended radio emission was not detected, an upper limit on $L_\textrm{ext}$  was estimated by taking the lowest frequency measurement for the core and extrapolating that to 300 MHz, assuming a lobe spectral index of $\alpha = 0.7$. 

%Upper limits were calculated for 1022 sources.

\subsection{SED Fitting}
\label{sec:sedfit}

\begin{figure}
    \centering
   \includegraphics[width=0.5\textwidth]{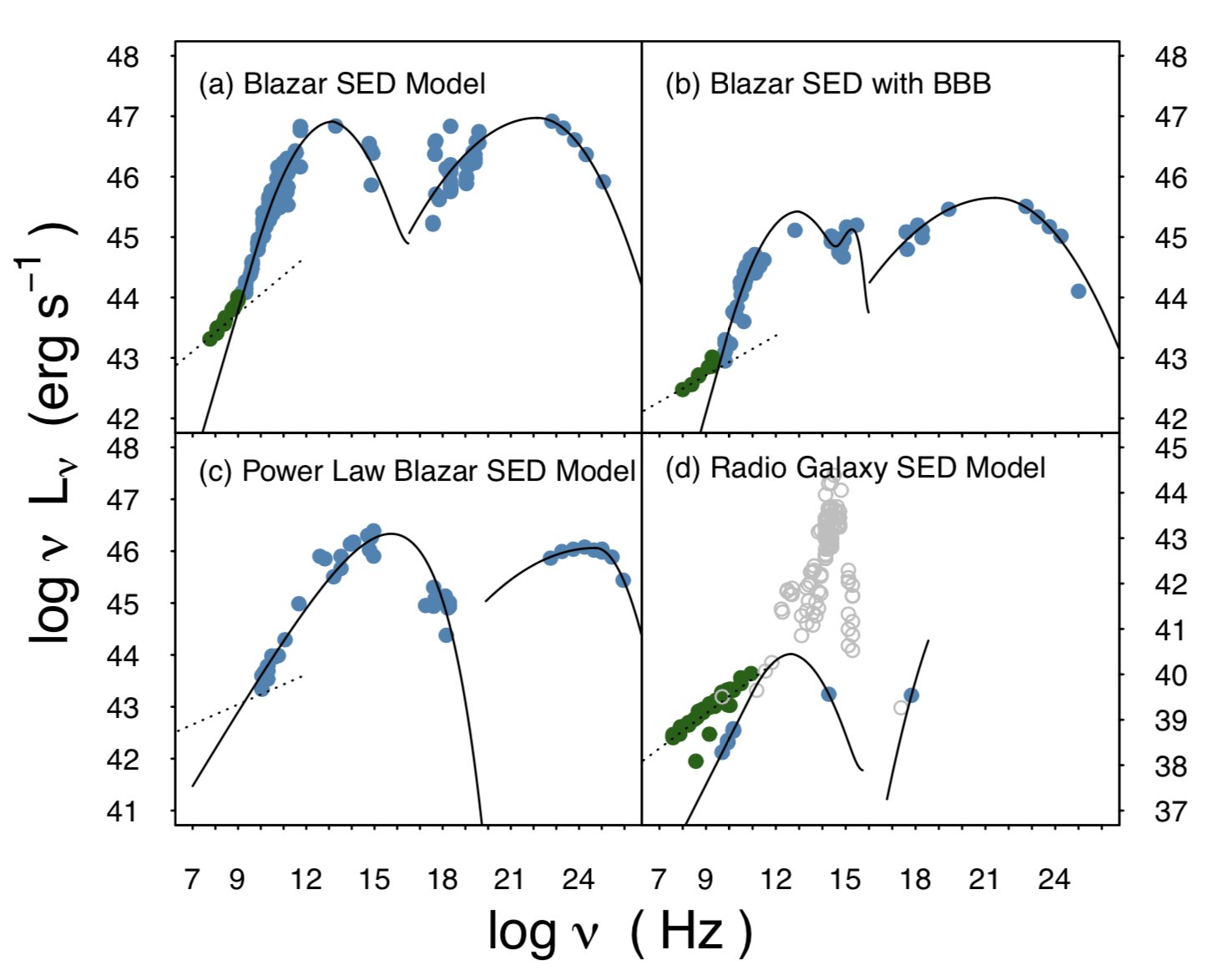}
    \caption{Example SEDs and model fits from our sample of jets.  All panels show the observations used in the fitting routine in blue, extended emission in green, and data not used in the fitting in gray.  The solid lines denote the resulting SED fits, and the dotted line shows the fits to the extended emission. Panel (a) shows an example of the parametric SED model of M11.  Panel (b) includes a thermal component due to the hot  accretion disk. Panel (c) shows an example of a high-energy peaked source, fit with a power law model with an exponential cutoff.  This source also only has an upper limit on the extended emission, extrapolated from the lowest-frequency data point.  Panel (d) shows a radio galaxy, for which only clearly identified non-thermal jet fluxes were used.}
    \label{sed_fig}
\end{figure}

%The broadband (radio to gamma-ray) jet SED is typically comprised of a core synchrotron component peaking between IR and X-ray energies, as well as a component at higher energies usually attributed to inverse Compton emission. 
For sources identified as blazars (or otherwise clearly dominated by the non-thermal jet), the synchrotron portion of the spectrum was fit with the parametric SED model from M11 in order to estimate $\nu_\textrm{peak}$ and $L_\textrm{peak}$. An example is shown in panel (a) of Figure \ref{sed_fig}, where the fluxes used in the fit are shown as blue data points and the fit as a black line.  If there was extended emission present, as determined by the previously described radio spectral decomposition, only data above $\nu_\textrm{cross}$ was used in the fit, and data identified as from the extended radio emission is shown in green.  In cases where there the thermal emission from the accretion disk is visible (i.e., a `big blue bump' or BBB), a disk component was added to the model. All SEDs are visually examined for this additional component, which was added to the model when appropriate, as shown in panel (b) of Figure \ref{sed_fig}.  In cases with a high-peaking synchrotron spectrum ($\nu_\textrm{peak} > 10^{15}$ Hz), the SED was sometimes better represented with either a power law model with an exponential cutoff, shown in panel (c) of Figure \ref{sed_fig}, or a two-sided parabolic fit.  In this panel, we also show an example of a source in which no extended emission is detected either in maps or in decomposing the spectrum. The upper limit on $L_\textrm{ext}$ is found by extrapolating from the lowest-frequency data point as shown by the dashed line.  
%Finally, the high energy peak was fit with a double-sided log parabola to estimate the peak frequency ($\nu_\textrm{IC,p}$) and luminosity ($L_\textrm{IC,p}$). These IC peak fit values are included in the tabulated results for completeness; however we reserve our analysis and discussion of the IC-related properties of our sample to a future publication.

%maybe this is needed somewhere but it was awkward here
%In blazars, the core emission is generally variable and the data have not been taken simultaneously. We required well-sampled SEDs so that we could measure the average SED parameters.

Unlike in blazars, the SEDs of radio galaxies are not dominated by the jet, and thus emission from the host galaxy, molecular torus, and/or accretion disk will be clearly visible and typical dominate from far-IR to UV.  For radio galaxies, or any source with apparent non-jet contamination, we only used known nuclear fluxes for the jet SED fitting -- either derived by us as described in Section \ref{hst_fluxes} or from catalogs, or using fluxes from NED which were explicitly labeled `core,' `nuclear,' etc.  An example is in panel (d) of Figure \ref{sed_fig}, where the gray points show non-jet emission (open gray circles); these data points are excluded from the fit.  
%We also note that while an IC spectrum is shown, this example does not have a measured value for $L_\textrm{IC,p}$ or $\nu_{IC,p}$ due to lack of data.

In sources identified as radio galaxies where there was insufficient data density to fit the SED with the parametric model but published nuclear fluxes were available in the radio, optical/IR, and X-ray, we  used a multivariate kernel density estimator (KDE) to estimate the peak frequency, as described in M11.  A joint distribution of $R_\textrm{C}$, $\nu_\textrm{peak}$, the radio-optical spectral index $\alpha_\textrm{ro}$, and the optical-x-ray spectral index $\alpha_\textrm{ox}$ was calculated using data for all sources well-fit by the parametric SED model.  The KDE was then used to determine the most likely value of $\nu_\mathrm{peak}$ given the other values for the more sparsely sampled radio galaxy SEDs. The value of $L_\textrm{peak}$ was then estimated from $L_\textrm{core}$, based on the correlation between these two observables: $$\log L_\textrm{peak} = 0.597 \times (\log L_\textrm{core} - 40) + 44.156  \mathrm{(erg/s).}$$  A figure showing the correlation and fit is given in online Supplemental Figure~\ref{fig:supp_lcorelpeak}.

%\begin{figure}
%    \centering
%    \includegraphics[width=\linewidth]{lpeak_lcore.pdf}
%    \caption{The correlation between $L_{core}$ and $L_{peak}$ which was used to estimate $L_{peak}$ in cases where there was insufficient data to fit the synchrotron spectrum. SUPPLEMENTAL ONLINE}
%    \label{lpeak_lcore}
%\end{figure}
 
\begin{figure*} 
    \centering
   \includegraphics[width=1\linewidth]{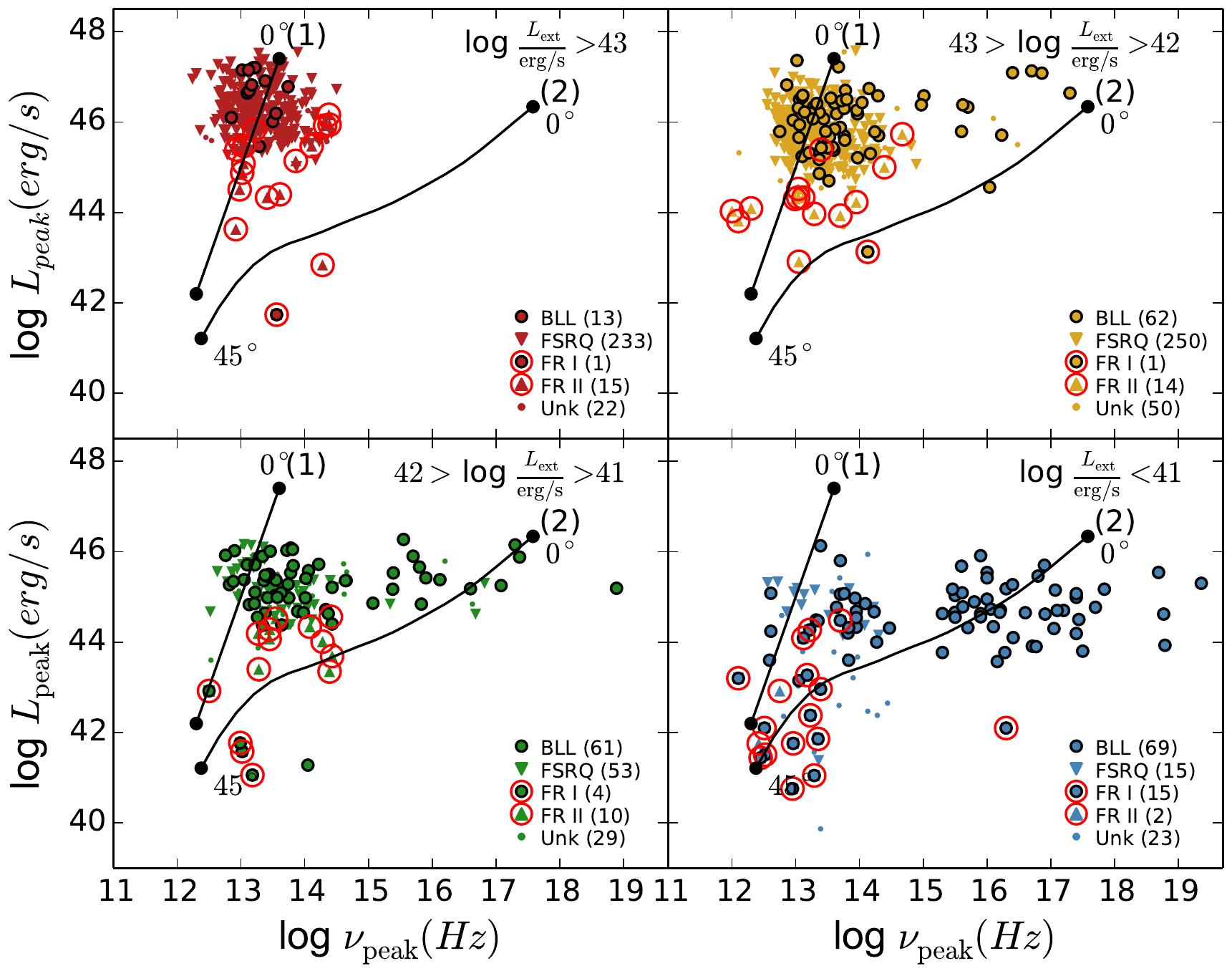}
    \caption{Synchrotron peak luminosity ($L_\mathrm{peak}$) versus peak frequency ($\nu_\mathrm{peak}$) for the TEX sample with measured extended radio luminosity ($L_\mathrm{ext}$), excluding those with missing redshift information. The sample has been divided into four broad bins in $L_\mathrm{ext}$ over the four panels, as labeled at top right. In all panels, BL Lacs and FR Is are shown as filled circles, FSRQs and FR IIs as triangles, and unknown object types as smaller dots, and the number of each type is given in each panel legend. Type I jets (BL Lac/FR I) have a black outline, while type II (FSRQ/FR II) do not.  All radio galaxies are enclosed in a red circle.  The plotted curves show the theoretical path of a jet through the \vplp\, plane as it is `debeamed' from 0$^\circ$ at the upper end of the track, to 45$^\circ$ at the opposite end. Track (1) corresponds to the expected debeaming path for a simple jet characterized by a single Lorentz factor, while track (2) is based on the decelerating flow model of Georganopoulos \& Kazanas (2003)}
    \label{4panel}
\end{figure*}
\subsection{The TEX and UEX Jet Samples}
\label{sec:final}
From the initial sample of nearly 7000 sources, we have selected those where the full SED had sufficient spectral coverage to reliably fit the synchrotron peak and return a value for the peak frequency and luminosity. All SEDs were assessed visually for goodness of fit, without any a priori knowledge about their identity or type, simply based on being well-fit with an appropriate amount of spectral coverage, following the same procedure as in M11. Generally, those that were eliminated lacked coverage over significant portions of the spectrum (e.g., no optical and/or X-ray) or could be reasonably fit by two or more very different SED shapes (often in these cases the X-ray spectral index is unknown -- see Appendix). 
%All SEDs are available as online supplemental figures.

A little more than one-third of the initial sample, or 2124 sources, remained after assessing the reliability of the broadband SED fit. Of these, 1045 sources have estimates of the extended radio luminosity, and %which are listed along with results of their SED analysis in Table \ref{sourcedata}. 
we call this sample the "Trusted Extended" (TEX) sample, as in M11. The TEX sample contains 575 FSRQs, 246 BL Lacs, 21 FR Is, and 41 FR IIs, and 162 sources of unknown object type. Another 1079 sources with good broadband SEDs only have upper limits on $L_\textrm{ext}$. We call this the "Unknown Extended" (UEX) sample as in M11.
%, and the fully-described table is available in the online supplementary material.  
It contains 688 FSRQs, 242 BL Lacs, 0 FR Is, 1 FR IIs, and 148 sources of unknown type.   All object types are based on either the literature reference giving published nuclear fluxes (for radio galaxies, listed in Table \ref{imported_data}) or from SIMBAD.  All redshifts ($z$) are taken from SIMBAD and/or NED. In the TEX sample 103 sources (169 in the UEX sample) have no redshift information available on SIMBAD/NED. In these cases a value of $z=0.3$ is assumed ($\approx 10\%$ of the TEX sample, 16\% of the UEX sample).  This value was chosen as it is roughly the transition from a nearby source to high-redshift source \citep[e.g.][]{birzan20}; however, none of our conclusions depend on this assumption, and we exclude these sources from plots with quantities (i.e., luminosities) that are strongly affected by the unknown redshift. A further discusson of these sources and their properties and location in the \vplp\, plane appears in the Appendix. %\textbf{Mary is this correct? you are not using NED?}. 

The properties of the TEX sample are given in Tables \ref{tex_radio} and \ref{tex_bb}. These include measurements of the black hole mass ($M_{BH}$) and apparent jet speed ($\beta_\mathrm{app}$) which are described in the next sections.  The radio properties of the TEX sample are given in Table~\ref{tex_radio}, which includes the source name in column 1, RA in column 2, DEC in column 3, redshift in column 4,  the 300-MHz extended luminosity ($L_\textrm{ext}$) in column 5, the 1.4 GHz core luminosity ($L_\textrm{core}$) in column 6, $\nu_\textrm{cross}$ in column 7, the radio core dominance ($R_c$) in column 8, apparent jet speed ($\beta_\mathrm{app}$) in column 9 and the reference for $\beta_\mathrm{app}$ in column 10.  Additional properties of the TEX sample are given in Table~\ref{tex_bb}, which includes the source name in column 1, object type (BLL, FSRQ, etc) in column 2, $L_\textrm{peak}$ in column 3, $\nu_\textrm{peak}$ in column 4, $M_\textrm{BH}$ in column 5 and the reference for $M_\textrm{BH}$ in column 6. Column 7 gives the original sample ID with a letter code corresponding to Table~1. 

The properties of the UEX sample are available in online Supplemental Table~\ref{supp:uex_tab1}.

\subsection{Black Hole Mass Measurements}
The analysis presented here relies on estimates of the central black hole mass for the jets in our sample. 
There are many different methods of measuring the black hole mass ($M_\textrm{BH}$) in AGN.  Reverberation mapping, generally considered the most reliable method, uses the  time delay between variations in the optical continuum and broad lines to give an estimate of the radius of the broad line region \citep[BLR;][]{2015arXiv150504805B}.  The black hole mass is then derived from this time delay and the velocity of the gas in the BLR \citep[as measured from the width of the lines,][]{1982ApJ...255..419B}.  Other methods take advantage of the observed correlations between $M_\textrm{BH}$ and the galaxy bulge luminosity \citep[which is an indicator of stellar mass, e.g.][]{marconi2003} as well as the correlation with the velocity dispersion of the stars in the galaxy \citep{ferrarese2000}.  Single-epoch spectroscopy takes advantage of scaling relations between $M_{BH}$ and emission line luminosities (which are typically from large surveys, such as the Sloan Digital Sky Survey).  This method only requires a single observation to estimate $M_{BH}$, making it perhaps the most accessible method for large samples of AGN \citep[e.g.][]{2017ApJ...850...74K}.  

We have attempted to exhaust the literature to compile $M_\textrm{BH}$ measurements for as many sources in the final sample as possible. Estimates of $M_\textrm{BH}$ are available for 227 sources in the TEX sample (274 sources for the combined TEX + UEX samples). %These are located in Table \ref{sourcedata}.  All values are given in units of $\log M_\odot$.  
In cases where there are multiple reported values of $M_\textrm{BH}$, we take the average.

\subsection{Apparent Jet Speed Measurements}
\label{sec:speeds}
%This analysis also uses measured apparent speeds as they can put interesting constraints on the Lorentz factor of the bulk flow within the jet as well as on the orientation of the jet. 
With Very Long Baseline Interferometry (VLBI), it is possible to resolve small features in the flow near the base of the jet \citep[e.g., ][]{boccardi17}, which can be traced between observations to measure their proper motions, or apparent angular speeds ($\mu$), which are typically on the order of mas/yr \citep[e.g.,][]{2019ApJ...874...43L,jorstad17,2018ApJ...853...68P}.  When converted into the apparent speed ($\beta_\textrm{app}$) in units of the speed of light, sources are found to have superluminal speeds, up to about 80c, implying highly relativistic flows.
We have compiled a large catalog of all jets with proper motion measurements from published sources. 
%A full description of this catalog and the references is described in Keenan et al, (2021), in preparation. 
Proper-motions monitoring programs often yield detections of multiple parsec-scale components with different speeds \citep[e.g.][]{lister16}. This could be due to an accelerating flow, differences in the flow speed over time, slight changes in the orientation angle, or variation in the angle of ejection relative to the main jet direction \citep[e.g.][]{2019ApJ...874...43L}. These values typically tend to cluster around a characteristic speed. Differences can be on the order of the maximum speed \citep{2013AJ....146..120L}, although differences this large are rare.  %Typical dispersion are on the order of xx\% of the maximum speed measured (cite). 
For this study, where multiple $\beta_\textrm{app}$ values exist for a jet, we adopted the largest observed value. These are given in Tables~~\ref{tex_bb}~and~\ref{supp:uex_tab1} along with the appropriate reference. In all, 400 (289, 111) sources in the combined (TEX, UEX) catalog have a VLBI proper-motion measurement.

\begin{table*}
    \centering
    \caption {TEX Radio Properties} 
    \begin{tabular}{c|c|c|c|c|c|c|c|c|c}
        Source Name & RA & DEC & Redshift & $L_\textrm{ext}$ & $L_\textrm{core}$ & $\nu_\textrm{cross}$ & $R_c$ & $\beta_\mathrm{app}$ & Ref.\\
        -- & J2000 & J2000 & -- & erg/s & erg/s & Hz & -- & c & --\\
        (1)&(2)&(3)&(4)&(5)&(6)&(7)&(8)&(9)&(10)\\
        \hline
                WB 2359+1439 & 00 01 32.83 & +14 56 08.13 & 0.3988 & 42.06 & 41.38 &  10.056 &  -0.88 &     ... & \\
                PKS 2359-221 & 00 02 11.98 & -21 53 09.86 &    ... & 41.58 & 42.05 &   8.471 &   0.45 &     ... & \\
                 B3 0003+380 & 00 05 57.18 & +38 20 15.15 & 0.2290 & 40.82 & 41.87 &   8.303 &   1.12 &  5.0950 & c \\
                 1Jy 0003-06 & 00 06 13.89 & -06 23 35.34 & 0.3467 & 41.68 & 42.80 &   8.262 &   1.29 & 11.6540 & t \\
                 PKS 0007+10 & 00 10 31.01 & +10 58 29.51 & 0.0900 & 39.96 & 40.22 &   9.207 &  -0.09 &  1.6560 & o \\

        \hline \\
    \end{tabular}
    \label{tex_radio}
    \begin{tablenotes}
    \item Table \ref{tex_radio} is published in its entirety in the machine-readable format.  A portion is shown here for guidance regarding its form and content. $L_\textrm{ext}$, $L_\textrm{core}$, and $\nu_\textrm{cross}$ are $\log_{10}$ values. The letters in column (10) denote the following references for $\beta_\mathrm{app}$ measurements: 
    (a) \cite{2008A&A...484..119B}
    (b) \cite{2007AJ....133.2357P}
    (c) \cite{lister16}
    (d) \cite{2019ApJ...874...43L}
    (e) \cite{1994ApJ...430..467V}
    (f) \cite{2018ApJ...853...68P}
    (g) \cite{lister09}
    (h) \cite{2012ApJ...758...84P}
    (i) \cite{2013AJ....146..120L}
    (j)\cite{2004ApJ...609..539K}
    (k) \cite{2001ApJS..134..181J}
    (l) \cite{2011AJ....142...49S}
    (m) \cite{jorstad17}
    (n) \cite{2001ApJ...549..840H}
    (o) \cite{2015MNRAS.446.2921F}
    (p) \cite{2001AJ....122.2954P}
    (q) \cite{2006A&A...446...71S}
    (r) \cite{2017MNRAS.466..952A}
    (s) \cite{2016A&A...586A..60K}
    (t) \cite{2005AJ....130.1418J}
    (u) \cite{2002ApJ...577...69J}
    (v) \cite{2007ApJ...659..225G}
    (w) \cite{2002ApJ...579L..67E}
    (x) \cite{2012A&A...544A..89L}
    (y) \cite{2010A&A...511A..57B}
    (z) \cite{2016A&A...585A..33B}
    (aa) \cite{2004ApJ...600..115P}
    \end{tablenotes}
\end{table*}

\begin{table*}
    \centering
    \caption {TEX Broadband Properties} 
    \begin{tabular}{lcccccc}
        Source Name & Type & $L_\textrm{peak}$ & $\nu_\textrm{peak}$ & $M_\textrm{BH}$ & Ref.  & Sample IDs\\
        -- & -- & erg/s & Hz & $M_\odot$ & -- & --\\
        (1)&(2)&(3)&(4)&(5)&(6)&(7)\\
        \hline
                PKS 0000-006 &        BLL & 44.65 & 13.98 & ... &  & t\\
               IVS B0001-120 &        BLL & 44.53 & 12.70 & ... &  & f,h\\
                PKS 0002-170 &       FSRQ & 45.37 & 14.13 & ... &  & h,o\\         
                     NGC 315 &       FRII & 41.76 & 12.42 &  9.0 & p,n,i & h,m,w\\
                TXS 0110+495 &       FSRQ & 45.05 & 13.92 &  8.3 & p & e,f,h\\
        \hline \\
    \end{tabular}
    \label{tex_bb}
    \begin{tablenotes}
    \item Table \ref{tex_bb} is published in its entirety in the machine-readable format.  A portion is shown here for guidance regarding its form and content. All measurements in this table other than redshift are $\log_{10}$ values. The letters in column (6) denote the following references for $M_{BH}$ measurements: 
    (a) \cite{2003ApJ...583..134B} 
    (b) \cite{2015PASP..127...67B}
    (c) \cite{2016yCat..74543864B}
    (d) \cite{2009MNRAS.397.1713C}
    (e) \cite{2017ApJ...850...74K}
    (f) \cite{2017ApJS..228....9K}
    (g) \cite{2006ApJ...642..711L}
    (h) \cite{2006ApJ...637..669L}
    (i) \cite{2010MNRAS.407.2399M}
    (j) \cite{2005MNRAS.361..919P}
    (k) \cite{2011MNRAS.413..805P}
    (l) \cite{2017A&A...598A..51R}
    (m) \cite{2018A&A...614A.120S}
    (n) \cite{2016ApJ...831..134V}
    (o) \cite{2004ApJ...615L...9W}
    (p) \cite{2002ApJ...579..530W}
    (q) \cite{2005ApJ...631..762W}
    (r) \cite{2002A&A...389..742W}
    (s) \cite{2004A&A...424..793W}
    (t) \cite{2005AJ....130.2506X}
    
    \end{tablenotes}
\end{table*}

\section{Results and Discussion}
\label{sec:results}

\subsection{Jet Power and the synchrotron \vplp\, plane}

Plots of $L_\mathrm{peak}$ versus $\nu_\mathrm{peak}$ are shown in Figure~\ref{4panel} for the TEX sample (excluding those with missing redshift information), binned on $L_\mathrm{ext}$. In all panels, BL Lac and FR I jets are shown as large circles with black outlines, FSRQ and FR II are plotted as triangles, and sources of unknown type as small circles.  All radio galaxies are enclosed in a red circle.  
%To examine how jet power\footnote{We will use the terms extended radio power and jet power interchangeably since the latter can be derived from the former, as discussed in section~\ref{sec:lowfreq}.} relates to the source location in the \vplp ~plane, we divided the sample into four broad bins based on extended radio luminosity corresponding to the four panels shown. It is clear that source location in the plane is strongly influenced by the jet power. 
%Given the probable large uncertainties in its measurement, $L_\textrm{ext}$ is not considered very reliable for any single source, and with broad binning the middle bins on $L_\textrm{ext}$ are likely very mixed. However, the jets at the extremes ($L_\textrm{ext} > 10^{43}$ erg/s and $L_\textrm{ext} < 10^{41}$ erg/s) are unlikely to contain many sources belonging to the other sample (i.e. there are very few sources in the lowest-power bin that truly belong in the highest power bin). 

With now nearly ten times the number of sources as in the original \cite{fossati98} blazar sequence, we still see some hint of an anti-correlation, in that the most powerful jets (red sources in top left panel) cluster exclusively at low $\nu_\textrm{peak}$ ($<10^{15}$~Hz) and high $L_\mathrm{peak}$, while jets at lower powers are `allowed' to have values of $\nu_\textrm{peak}$ up to $\sim10^{19}$~Hz, and appear to have somewhat lower values of $L_\mathrm{peak}$ (note however that this statement will be revised in the next section when UEX sources are included). Meanwhile, practically all radio galaxies regardless of type or power cluster at low $\nu_\textrm{peak}$, as previously noted in M11. We plot over the sources in Figure~\ref{4panel} two example `de-beaming paths' showing how the synchrotron peak luminosity and frequency will change for a theoretical jet as it is misaligned (from the point labeled 0$^\circ$), which agree with this convergence of misaligned jets at lower left. Track (1) shows the expected rapid drop in luminosity (4:1 in log, compared to frequency) under the assumption of a `simple' jet with a single-velocity flow (characterized by a single Lorentz factor $\Gamma$). Such a path appears consistent with the relative location of the powerful FSRQ (red triangles at top left of Figure~\ref{4panel}) and their presumed powerful radio galaxy counterparts just below. In contrast, track (2) shows the effect of significant velocity structure in the jet, where we use the decelerating flow model of \cite{markos03}. In this case, as the jet is misaligned the slower layers of the jet (which have lower characteristic $\nu_\mathrm{peak}$) come to dominate, and as a result $\nu_\mathrm{peak}$ can decrease rapidly as the viewing angle of the jet increases, resulting in a more horizontal de-beaming track. This appears to match the distribution of the lowest-power sources in blue at bottom right of Figure~\ref{4panel}. (The discussion of velocity gradients in type I jets is resumed in Section~\ref{sec:gradients}.)

Looking at spectral types, we see that practically \emph{all} broad-lined sources (FSRQ and FR II) appear to have low peak frequencies, with an upper bound at $\nu_\mathrm{peak}\sim 10^{15}$ Hz. We propose to call these "type II" jets (quivalent to the "strong" jets of M11), and define the `strong zone' as that bounded by $L_\mathrm{peak} > 10^{45}$ erg/s and $\nu_\mathrm{peak}< 10^{15}$ Hz. We propose type I jets as those with intrinsically low-excitation optical spectra associated with low-efficiency accretion. In compliment to the strong zone, we define a `weak zone' of  $\nu_\mathrm{peak} > 10^{15}$ Hz  (see online supplemental Figure~\ref{fig:supp_boxes}); as this selects a population that is almost entirely lineless BL Lac objects. Assuming that jets do de-beam like the example paths shown, this definition of zones should avoid the mixture of jet types in the misaligned population at the bottom left of the \vplp\, plane. As we will show, the characteristics of the jets in these two zones differ greatly.

With the zones defined, we make three important observations. First, that there are a substantial population of BL Lacs in the strong zone. The BL Lacs at low power and lower $L_\mathrm{peak}$ may be misaligned versions of the more aligned BL Lacs in the weak zone proper (i.e, they are in the transition between blazars and radio galaxies). However, at the highest powers (i.e. BL Lacs shown as red circles at top left of Figure~\ref{4panel}), it is unlikely these are misaligned jets, as we would expect to see the aligned counterparts (e.g. there are no high-$\nu_\mathrm{peak}$ sources at these jet powers). We will return to the possibility that many of the low-peak-frequency BL Lacs in the strong-jet zone are actually misidentified FSRQ (and argue that these should be considered type II jets) in section~\ref{spectraltype}.

The second observation is that there are a sizable number of FSRQs with very low jet powers, seen as the blue filled triangles with $L_\mathrm{ext}<10^{41}$ erg/s at lower right in Figure~\ref{4panel}. If broad-lined sources are always type~II jets, this suggests that type I/II jets are not divided at a certain power, but rather that type~II jets are possible at \emph{any} jet power. This also matches recent discoveries of large numbers of FR II radio galaxies at low powers \citep{2017A&A...601A..81C,2019MNRAS.488.2701M,2020ApJS..247...53K}. 

The third observation is that for type I jets (those with $\nu_\mathrm{peak}>10^{15}$ Hz) there appears to be an upper bound on the allowed jet power, as noted by the lack of these sources in the first panel of Figure~\ref{4panel} (with $L_\mathrm{ext}>10^{43}$ erg/s). As we will argue in Section~\ref{accretion_divide}, type I sources likely correspond to inefficient accretion systems, and thus this upper bound could naturally arise from the observed upper limits on black hole mass ($10^9-10^{10}\,M_\odot$) and the upper bound on the Eddington ratio (e.g., about 0.5\% of Eddington) for the inefficient mode. The most powerful type I jets will be hosted by systems at these limits. In comparison, the efficiently accreting systems that host type II jets, with the same range of black hole masses, will be able to reach jet powers at least one order of magnitude higher.

\subsection{The End of the Blazar Sequence}
\label{sec:blaz_seq}
\begin{figure*}
    \centering
   \includegraphics[width=\linewidth]{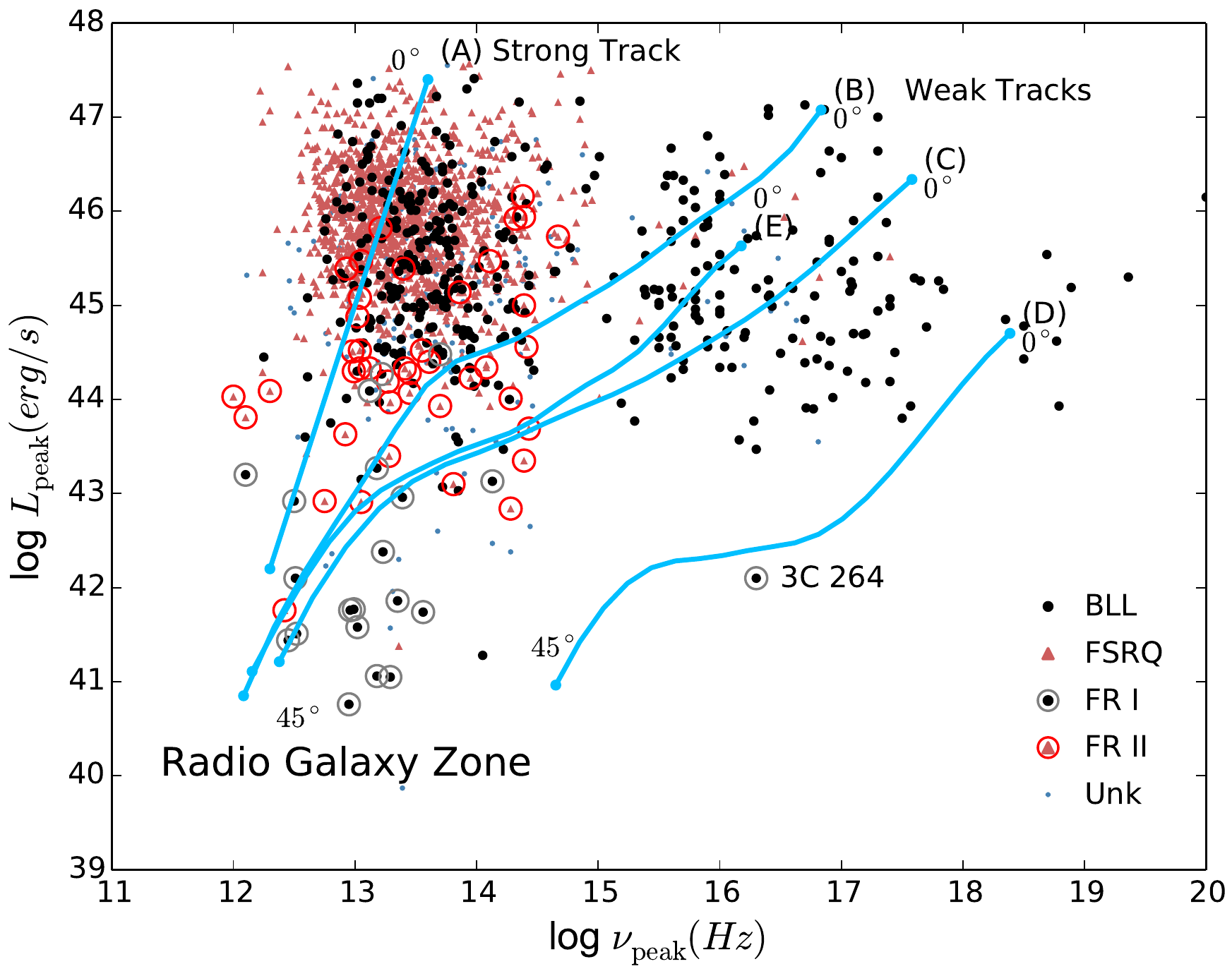}
    \caption{Synchrotron peak luminosity ($L_\textrm{peak}$) versus peak frequency ($\nu_\textrm{peak}$) for the TEX and UEX samples combined, excluding those with missing redshift information. Here BL Lacs and FR Is are plotted as black filled circles, and FSRQ and FR IIs as light red triangles. Radio galaxies are also circled in gray (FR I) or red (FR II). The source 3C 264 is labeled as it is a topic of discussion in the main text.  Paths (A), (B), (C), and (D) show theoretical de-beaming paths, as discussed in the main text.}
    \label{blazar_env}
\end{figure*}

In Figure~\ref{blazar_env} we display $L_\mathrm{peak}$ versus $\nu_\mathrm{peak}$ for the combined UEX and TEX samples (excluding those with missing redshift information). Here BL Lacs and FR I are shown as large black circles, FSRQ and FR II as red triangles, and unknown types as small blue points. Radio galaxies are further enclosed in circles, red for FR IIs and gray for FR Is. The theoretical de-beaming curves overlaid on the figure will be discussed in the next section.  It is immediately obvious that with the UEX sample included, there is a greater number of sources (nearly all BL Lacs) at high $\nu_\textrm{peak}$ and high $L_\textrm{peak}$ than in Figure \ref{4panel}. In the combined sample there are 30 sources with $\nu_\mathrm{peak} > 10^{15}$ Hz and $L_\mathrm{peak} > 10^{46}$ erg/s while in the TEX there are only 11.  A handful of sources in or near this region have previously been described \citep[e.g.][]{2003ApJ...588..128P,2006A&A...445..441N,2012MNRAS.422L..48P,2012A&A...541A.160G,cerruti2017}, and our results appear to agree with previous work showing that there is no `forbidden zone' at upper right in the \vplp\, plane. 

\begin{figure}
    \centering
    \includegraphics[width=\linewidth]{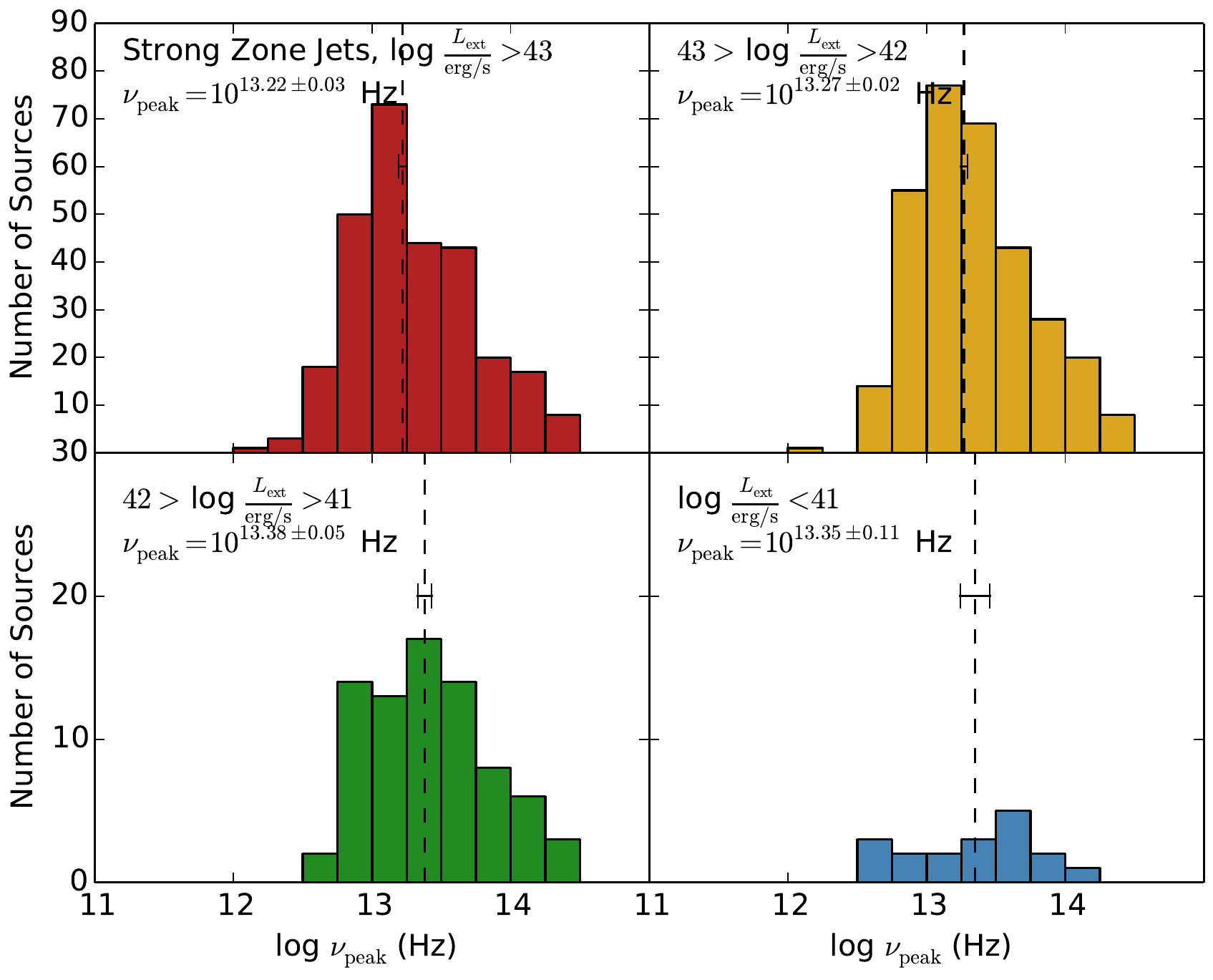}
    \caption{Histograms showing the distributions of $\nu_\textrm{peak}$ for sources in the strong/type~II zone,  binned on $L_\mathrm{ext}$ (range noted in the upper right corner).  The average values are shown as dashed vertical lines, with errors displayed as the horizontal bars. The average values of  $\nu_\textrm{peak}$ are found to remain relatively constant with increasing $L_\mathrm{ext}$. }
    \label{fig:fsrq_nupeak}
\end{figure}

\begin{figure}
    \centering
    \includegraphics[width=\linewidth]{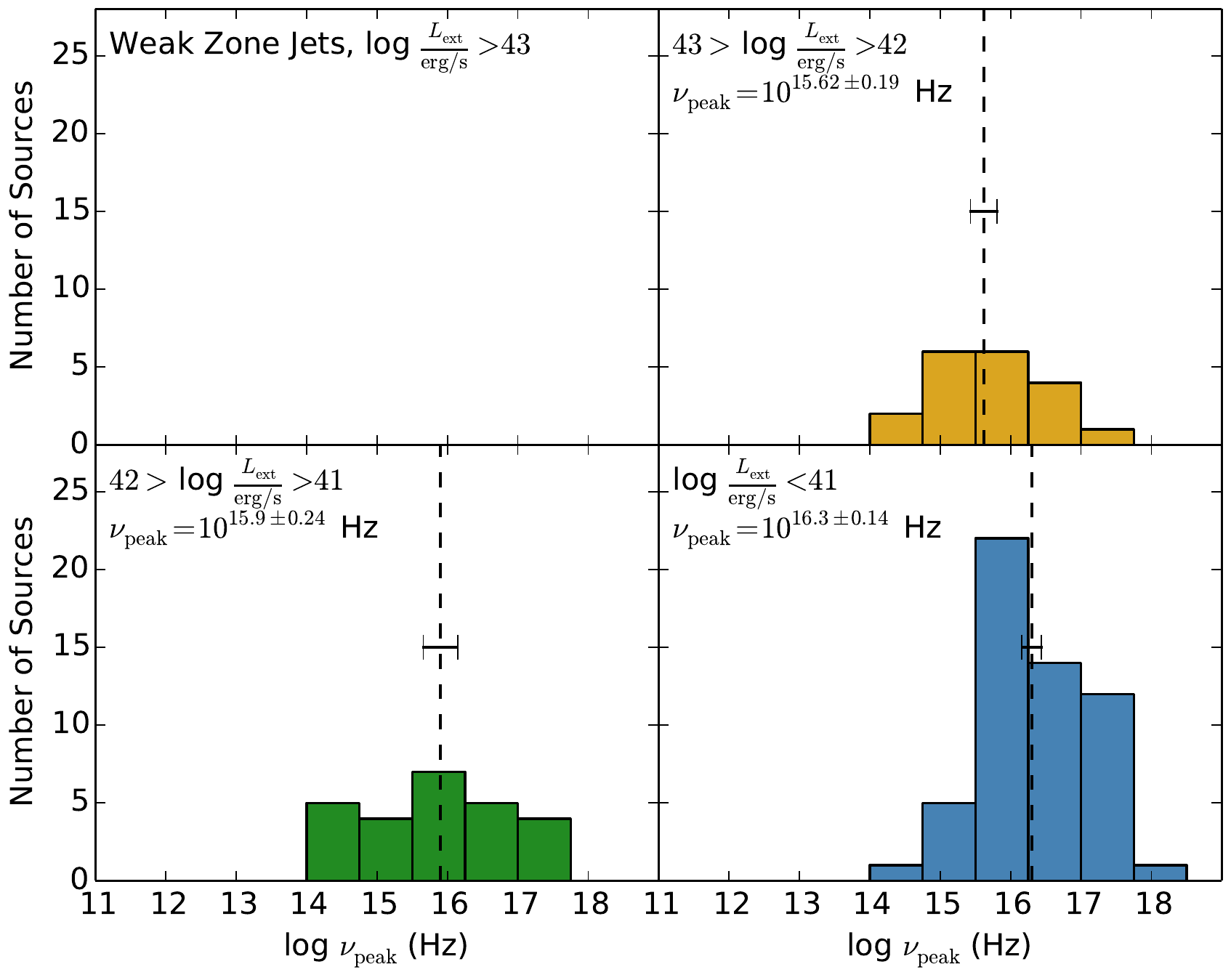}
    \caption{Histograms showing the distributions of $\nu_\textrm{peak}$ for weak-zone BLLs ($\nu_\textrm{peak} > 10^{15}$ Hz), binned on $L_\mathrm{ext}$ (as shown in the upper right corner).  The average values are shown as dashed vertical lines, with errors displayed as the horizontal bars.  The average values of $\nu_\textrm{peak}$ for these sources are found to remain constant with increasing $L_\mathrm{ext}$. Note: there are no sources in the highest $L_\mathrm{ext}$ bin.}
    \label{fig:bll_nupeak}
\end{figure}
Sources with high $\nu_\mathrm{peak}$ and high $L_\mathrm{peak}$ appear contrary to the original blazar sequence, which implied an anti-correlation between power and peak frequency, and more importantly, suggested that jets are mono-parametric.  A deeper investigation of these particular sources in our catalog is deferred to future work, but we can make some observations from the data available.  First, we can note that none of the forbidden-zone sources in our sample lack redshifts, and so their presence in this zone is not due to over-estimating the luminosity through an incorrectly assumed redshift.\footnote{Plots of $L_\mathrm{peak}$ versus $\nu_\mathrm{peak}$ with a color scale on redshift for the full sample, and with a color scale on $L_\mathrm{ext}$ for the UEX sample (excluding those with missing redshift information), are given in online supplemental Figures~\ref{fig:supp_redshift}~and~\ref{fig:supp_uex_uls}}. We further find that none of the UEX sources have an upper limit on  log~$L_\mathrm{ext}$ above $10^{43}$ erg/s. Thus a possible explanation for the high-$\nu_\mathrm{peak}$, high-$L_\mathrm{peak}$ sources preferentially appearing in the UEX sample is that these are highly aligned type I jets with relatively low $L_\mathrm{ext}$. In such cases, the core spectrum will dominate down to very low radio frequencies making an estimate of $L_\textrm{ext}$ difficult given the large dynamic range between the highly beamed core and low-luminosity extended emission (this is further supported by the range of radio core dominance discussed in Section \ref{sec:orientation}).

In previous work, some of the authors of this paper suggested that although the type I/II (weak/strong) jet divide and jet velocity gradients alters the original idea of the blazar sequence, there may still be `sequence-like' behavior within the two sub-populations, such that location in the \vplp~ plane would be determined by jet power \citep{2011arXiv1111.4711G,2012arXiv1205.0794M}. The most straightforward test of this is to simply look at how type I and II jets behave in terms of the synchrotron peak frequency as a function of $L_\mathrm{ext}$.

As previously noted, the FSRQs in our sample range from the highest-power sources down to very low $L_\mathrm{ext}$ %($<10^{41}$ erg/s), 
which appear to match recent observations of low-luminosity sources with FR II morphologies  \citep{2019MNRAS.488.2701M} as well as the jets from smaller ($10^6-10^7\,M_\odot$) black holes seen in narrow-line Seyfert galaxies \citep[e.g.][]{abdo2009_nlsy,foschini2017}. It is possible that the `downsized' low-power type~II jets in general come from smaller black holes (see Section~\ref{accretion_divide}). As shown in Figure~\ref{fig:fsrq_nupeak}, we see that type~II jets have a remarkably consistent distribution of synchroton peak frequencies over 4 orders of magnitude in jet power, as every bin has an average value of approximately $\nu_\mathrm{peak} = 10^{13.3}$ Hz (the average values and errors are obtained from a Gaussian fit, with exact values noted in the figure).  \cite{ghisellini2017}, in a study of blazars detected by \emph{Fermi}/LAT, similarly found that FSRQ do not appear to follow the blazar sequence.

We also find no significant difference between the average $\nu_\mathrm{peak}$ value for sources above  $\nu_\mathrm{peak}$=$10^{14.5}$\,Hz (Figure~\ref{fig:bll_nupeak}), though the distributions do appear to differ, with the lowest-power type I jets having a larger number of extremely high-$\nu_\mathrm{peak}$ sources, and the highest-power sources seeming to prefer lower $\nu_\mathrm{peak}$ values.  For type I jets, the effects of a horizontal de-beaming path through the \vplp~plane and the small number of sources may make it difficult to discern a real trend. On the whole the evidence does not strongly support `blazar sequence-like' behavior within either jet population. 

%, and argue that these are low-power sources which live in very sparse environments, allowing the jets to remain undisrupted and terminate in hotspots. This shows that the jet power and the large-scale morphology are not as connected as previously thought. Others have found similar sources \citep{2017A&A...601A..81C,2020ApJS..247...53K}. There should be a matching population of misaligned jets, i.e., low-power FR IIs. In our sample there are only 2 FR IIs at $L_\mathrm{ext}<10^{41}$ erg/s. These sources may still be characterized by relatively fast flows with a simple debeaming pattern, so that with low intrinsic powers, the observed flux of misaligned jets quickly drops to become very faint, thus rarely make it into flux-limited surveys \citep{2019MNRAS.488.2701M}. 
%In order to detect low-power FR IIs, they would have to be relatively nearby, and since they exhibit positive cosmological evolution there are very few nearby \citep{schmidt68,dunlop_peacock,2012ApJ...751..108A}, thus very few exist in the TEX sample. 

\subsection{Velocity Gradients in Type I Jets}
\label{sec:gradients}
Returning to Figure~\ref{blazar_env}, we now discuss the theoretical curves for the location of a jet in the \vplp\, plane as a function of orientation angle. The curves all begin with fully aligned jet at 0$^\circ$ at upper right, and show how the jet moves through the plane as it is misaligned down to $45^\circ$.  Track A shows the expected path for a single-velocity flow, as described for Figure~\ref{4panel}, while tracks B, C, D, and E show example de-beaming paths for jets with velocity gradients using the decelerating flow model of  \cite{markos03} with different parameters. In all cases, the jet has a characteristic initial and final Lorentz factor ($\Gamma_\textrm{init}, \Gamma_\textrm{fin}$), and the deceleration occurs over a certain length scale \cite[$L_\mathrm{decel}$, see][for further details of the model]{markos03}. The values for the different tracks are given in Table~\ref{table:decel_flows}. We note that these tracks are meant to be illustrative of structured jets in a general way.  We are not suggesting that all weak jets have decelerating flows in particular \citep[versus, e.g., a spine-sheath model,][]{ghisellini2005}. We would expect similar tracks in other types of structured jets and an extensive investigation of different theoretical models is beyond the scope of this work.

As shown, a simple jet model with velocity gradients allows us to explain the wide range of peak frequencies and luminosities of blazars in the weak-jet region with very reasonable values of $\Gamma$, while remaining consistent with the location of misaligned type I jets (FR Is) at lower left. Track (C) is the same as the example track in M11, shown here for reference. The parameters of track (B) have been chosen to explain the high-$L_\mathrm{peak}$, high-$\nu_\mathrm{peak}$ sources that contradict the original blazar sequence.  If our interpretation is correct, we should see signs that these high-frequency, high-luminosity sources are very well aligned (see Section \ref{sec:orientation}). 

The parameters of track (D) were chosen for the curve to pass near the few sources with relatively high $\nu_\textrm{peak}$ and low $L_\textrm{peak}$, one of which is the well-known source 3C 264, which hosts a superluminal optical jet \cite[][labeled in Figure \ref{blazar_env}]{2015Natur.521..495M}. The general lack of sources in this region is important evidence for velocity gradients in type I jets, since `simple' (single-velocity flow) de-beaming of the high-peak-frequeency BL Lacs should lead to a population of radio galaxies in this region. Although 3C~264 is nominally a radio galaxy, it is close to the `transition zone' between radio galaxies and blazars with an orientation angle at or below 10 degrees \citep{3c264_veritas_2020}. Thus it seems more likely that 3C~264 is a slightly misaligned low-power analog to the higher-power jets that live above it in the \vplp\, plane. The source is well-studied enough for inclusion likely only because it is very nearby ($z = 0.02$; $d_A$=91\,Mpc) and has a prominent optical jet. It is found in both the 3C and 4C surveys; a theoretical source at the same radio luminosity would drop out of the 3C survey beyond $z \approx 0.03$, and from the 4C survey beyond $z \approx 0.08$. Thus similar sources likely exist, but would require dedicated deep multi-wavelength surveys to detect and characterize. 

Based on the idea that something like de-beaming track (C) could describe most of the sources in the type I-jet half of the \vplp\, plane, M11 suggested that the most aligned type I jets are those with the very highest values of $\nu_\mathrm{peak}$, and that intermediate-peak BL Lacs (IBLs\footnote{Here we note the common usage of "low-peak", "intermediate-peak" and "high-peak" BL Lacs, or LBLs, IBLs, and HBLs. The exact dividing line between classes varies slightly in the literature, but we adopt $10^{14.5} \mathrm{Hz}<\nu_\mathrm{peak}<10^{15.5}$ Hz for IBLs with HBLs and LBLs above and below this range, respectively.}) would thus be misaligned counterparts of the former according to the velocity gradient de-beaming path. However, it is possible that the IBL population also includes `intrinsic' IBLs which have intermediate peak frequencies when fully aligned. This is highlighted by track (E), which shows an example of a jet which begins at 0 degrees near the IBL zone.  The possibility of a mixture of orientations in the IBL zone is further discussed in the next section.

\begin{table}
    \centering
    \begin{tabular}{c|c|c|c}
         Track & $\Gamma_\mathrm{init}$ & $\Gamma_\mathrm{fin}$ & $L_\mathrm{decel}$  \\
         -- & -- & -- & cm \\
         \hline
         B & 26 & 3 & $1\times 10^{17}$\\
         C & 15 & 3 & $5\times 10^{16}$\\
         D & 6 & 3 & $2\times 10^{16}$\\
         E & 10 & 3 & $1\times 10^{17}$
    \end{tabular}
    \caption{The parameters describing the decelerating flow tracks in Figure \ref{blazar_env}.}
    \label{table:decel_flows}
\end{table}

%%%%%%%%%%%%%%%%%%%%%%%%%%%%%%%%%%%%%%%%%%55
% DONT DELETE: May want to bring this in for the conclusions, or some part of it

%It is important to note that while velocity gradients do allow the horizontal movement through the \vplp\, plane, it is not the velocity gradients per se which explain the fact that lower-power sources are `allowed' to go to high $\nu_\textrm{peak}$ while high-power sources are not.  The characteristic peak frequency is intrinsic to the underlying radiating particle distribution.  The structured jets likely have a range in their Lorentz factors ($\Gamma$).  The high energy emission might originate in the fast-moving parts which do have high $\Gamma$ (e.g. the spine in a spine-sheath model, or the base of the jet in a decelerating flow model).  Strong jets likely also have high $\Gamma$, however, something is preventing these sources from having high $\nu_mathrm{peak}$.  Strong sources are thought to have large external photon fields [NOTE THAT THE EXTERNAL FIELD ENERGY DENSITY IS THOUGHT TO BE FIXED], which could promote the cooling of high energy radiation, shifting the peak to lower frequencies \citep{ghisellini98}.  In contrast, weak sources are thought to have radiatively inefficient accretion disks, which may not be able to sustain a broad line region or a dusty torus \cite{2009ApJ...701L..91E}, resulting in smaller photon fields. The lack of these photons allows the peak frequency to remain at high energies.  

\subsection{Orientation}
\label{sec:orientation}

\begin{figure}
    \centering
   \includegraphics[width=1\linewidth]{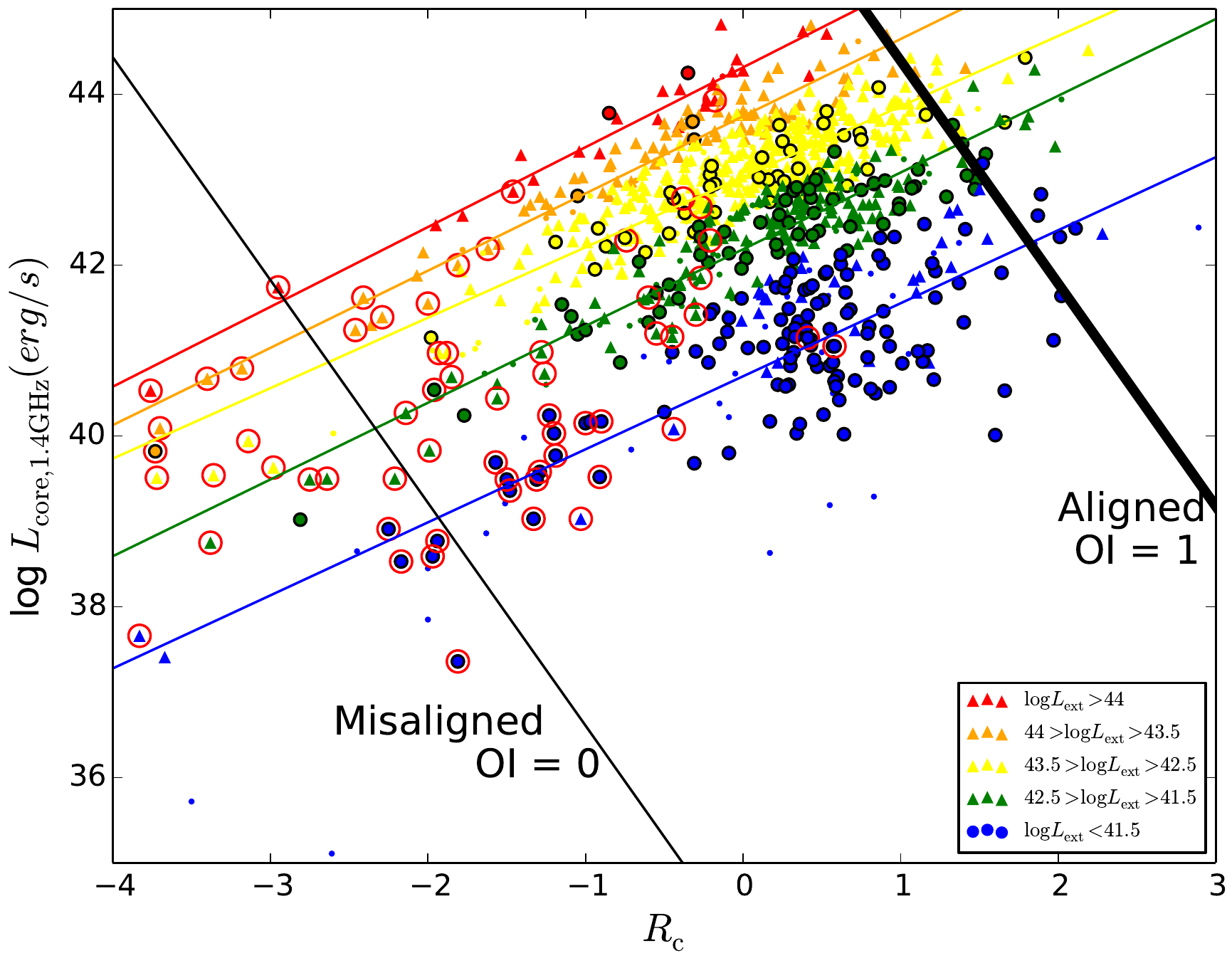}
    \caption{Radio core luminosity at 1.4 GHz ($L_\textrm{core}$) versus the radio core dominance at 1.4 GHz ($R_\textrm{c}$) for the TEX sample, excluding sources with missing redshift information. The color scale is on extended radio power ($L_\textrm{ext}$) as sown in the legend, with red sources being the most powerful.  The bins all show similar correlations (where we give the linear regression as respectively colored lines). We define a parametric orientation indicator (OI) based on where each source lies with respect to the black lines bounding this region (as described in the text).}
    \label{fig:coredom}
\end{figure}

\begin{figure}
    \centering
   \includegraphics[width=0.45\textwidth]{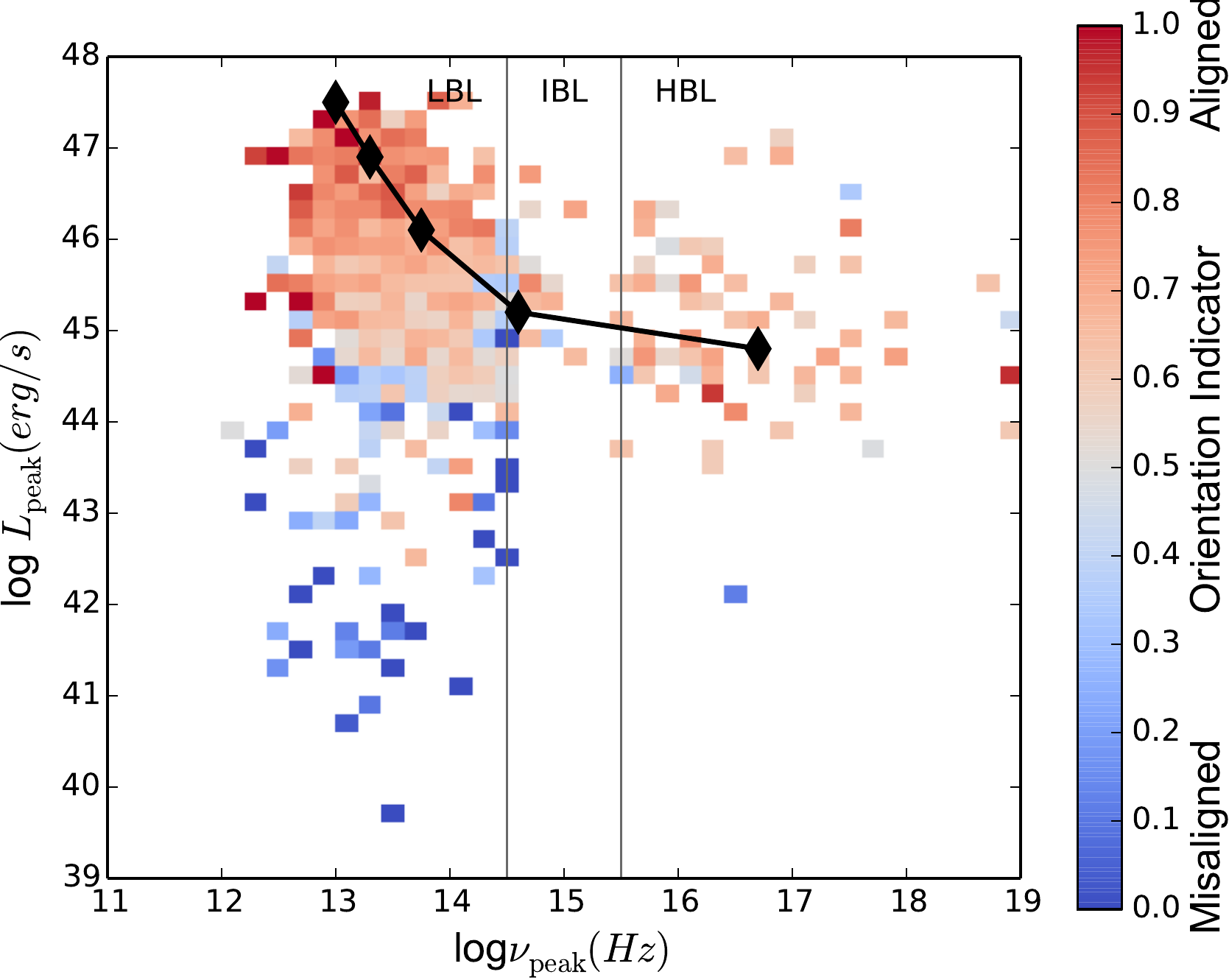}
    \caption{The \vplp~plane with color scale based on the new orientation indicator (OI) where a value of 1 corresponds to a highly aligned jet and a value of 0 to a misaligned jet. As expected, there is a noticeable gradient from aligned to misaligned as peak luminosity decreases. Their are indications that sources with intermediate synchrotron peak frequencies are more misaligned. The original blazar sequence of Fossati et al., (1998) is denoted by the connected black diamonds.}
    \label{fig:orientation_map}
\end{figure}
%While $L_\textrm{ext}$ can reasonably constrain the jet power,   

We have so far only used the crude distinction between blazar and radio galaxy to argue about the role of angle in filling the \vplp ~plane, as the orientation angle is difficult to measure precisely and is only well-constrained in a handful of cases. 
While radio core dominance ($R_\mathrm{c}$) has frequently been used as an orientation indicator \citep[e.g. ][]{1994ApJ...430..467V,1995ApJ...448L..81W,meyer2011,2016ApJ...830...82M}, it has not always lead to clear results \citep{2004MNRAS.348..937C}. 
%1995ApJ...448L..81W - suggest using optical/core luminosity instead of lobe/core -- but both quantities are beamed? 
This may be because the range of $R_\mathrm{c}$  is different depending on the jet power of the source. Indeed, the distribution of $R_\textrm{c}$ for the highest ($L_\mathrm{ext}>43.5$) and lowest ($L_\mathrm{ext}< 41.5$) power sources shows a decided shift (see histograms in online Figure~\ref{supp:hist_rce}). The bulk of highest-power sources peak around $R_\mathrm{c}\sim0$ and almost none have $R_\mathrm{c}>1)$. The lowest-power sources peak at $R_\mathrm{c}\sim$0.5-1 and extend up to much higher values (core dominance of 2 or 3 is not unusual). The apparent ceiling on maximum $R_\mathrm{c}$ for high-power jets is unlikely to be a selection effect against aligned sources: indeed, given a flux-limited sample high-power, highly aligned sources will be over, not under-represented, due to beaming effects.

In Figure \ref{fig:coredom}, we plot $R_\textrm{c}$ versus
the core luminosity at 1.4 GHz ($L_\textrm{core}$), with color representing $L_\textrm{ext}$ for the TEX sample (excluding those with missing redshift information). A thick black line is drawn across the plot in the region of the most aligned sources. Here again we note that the maximum radio core dominance appears to depend on the jet power, with low-power sources having higher $R_c$ when fully aligned. For example, a very high-power jet (red sources in Figure \ref{fig:coredom}) with $R_c\approx 0.5$ could be nearly fully aligned, while a very low-power source (blue sources in Figure \ref{fig:coredom}) with the same $R_c$ is well away from the most aligned sources of that power and in fact closer to the region of low-power radio galaxies (circled sources).

To roughly parameterize orientation, we assigned a value from 0 to 1 for each source relative to where the source lies with respect to the parallel black lines drawn in Figure~\ref{fig:coredom}; we call this the "orientation indicator" (OI).  An OI of 0 represents a source which is closer to the plane of the sky (i.e. at 90 degrees) and a value of 1 represents a source which is pointed along the line of sight. To better discriminate between the bulk of sources, the lines are not absolutely outside the entire population; outliers (sources outside of the black lines) were simply truncated to a value of either 0 or 1 as appropriate. The equation giving the OI value (before truncation) is $$\mathrm{OI} = 0.077 L_\mathrm{core}  + 0.201 R_c - 2.615$$ where $L_\mathrm{core}$ is the $\nu L_\nu$ power measured at 1.4 GHz and $R_c$ is the log ratio of the core to extended at 1.4 GHz.

A binned version of the \vplp ~plane for the TEX sample, with a color scale corresponding to OI is shown in Figure \ref{fig:orientation_map}.  This plot was created by separating the plane into 2D bins over $\nu_\mathrm{peak}$ and $L_\mathrm{peak}$, and taking the average value for the OI in each bin.  Red indicates that most of the sources which lie in that respective bin are well-aligned, while blue indicates misaligned, according to the scale shown at right.  We find that the along the original blazar sequence (denoted by the connected black diamonds), most sources are well-aligned (red), as expected.  We also find that for the type II (strong zone) jets, sources become gradually more misaligned (blue) as $L_\mathrm{peak}$ decreases. 
The sources in the weak zone more complicated, likely due to a combination of the smaller sample size as well as the mixed population at intermediate frequencies (the possibility of `intrinsic' IBLs as mentioned earlier).  
However it does seem that sources in the IBL region are more misaligned than those at the highest peak frequencies. It is not clear if these more misaligned jets are counterparts to the HBLs to the right in the plane or perhaps misaligned version of the high-frequency, high-luminosity sources which are largely absent from the TEX sample, as discussed previously. In general, the most misaligned sources are found in the lower left region of the \vplp~ plane where radio galaxies reside, as expected.  Our results support the general picture where structured (weak/type I) jets follow largely horizontal movement through the \vplp~plane. 

As a final comment on Figure~\ref{fig:coredom}, the fact that $R_c$ is highest in well-aligned, low-power jets may mean that these sources are less likely to have detections of their extended radio emission. To detect faint extended emission around a highly dominant core of 100-1000 times the extended flux would require radio imaging which is highly sensitive and with high dynamic range, which is challenging. This may explain why in general there are fewer low-power sources with known $L_\mathrm{ext}$ compared to the most powerful FSRQ. More particularly, it may explain the population of high-frequency, high-luminosity sources in the UEX sample which seem to be largely missing from the TEX sample, as previously discussed in Section~\ref{sec:blaz_seq}. Future studies with deep, high-dynamic-range radio imaging of these sources can confirm this.  We also note that our definition of OI is far from rigorous and could be affected by selection effects, and the use of it here is meant to be more illustrative than conclusive. It is left to future work with more statistically complete samples to examine this in more detail.

Another way to look at orientation is through measurements of the apparent jet speed ($\beta_\mathrm{app}$, parameterized in units of $c$), which is derived from observations of proper motions and has been measured for a large numbers of jets using very long-baseline radio interferometry \citep[VLBI, e.g.,][]{lister2019}. As $\beta_\mathrm{app}$ is a function of the bulk Lorentz factor ($\Gamma$) and the orientation $\theta$, it can be used to put constraints on both. In particular, the maximum angle that a jet can have can be found by letting the true jet speed $\beta$ equal unity. Although at very small orientation angles the apparent jet speed drops to zero because of extreme foreshortening, such highly aligned sources are rare, and for a jet population over a range of orientations with similar intrinsic speeds, we would expect more aligned jets to have faster observed $\beta_\mathrm{app}$. 

\begin{figure}
    \centering
   \includegraphics[width=\linewidth]{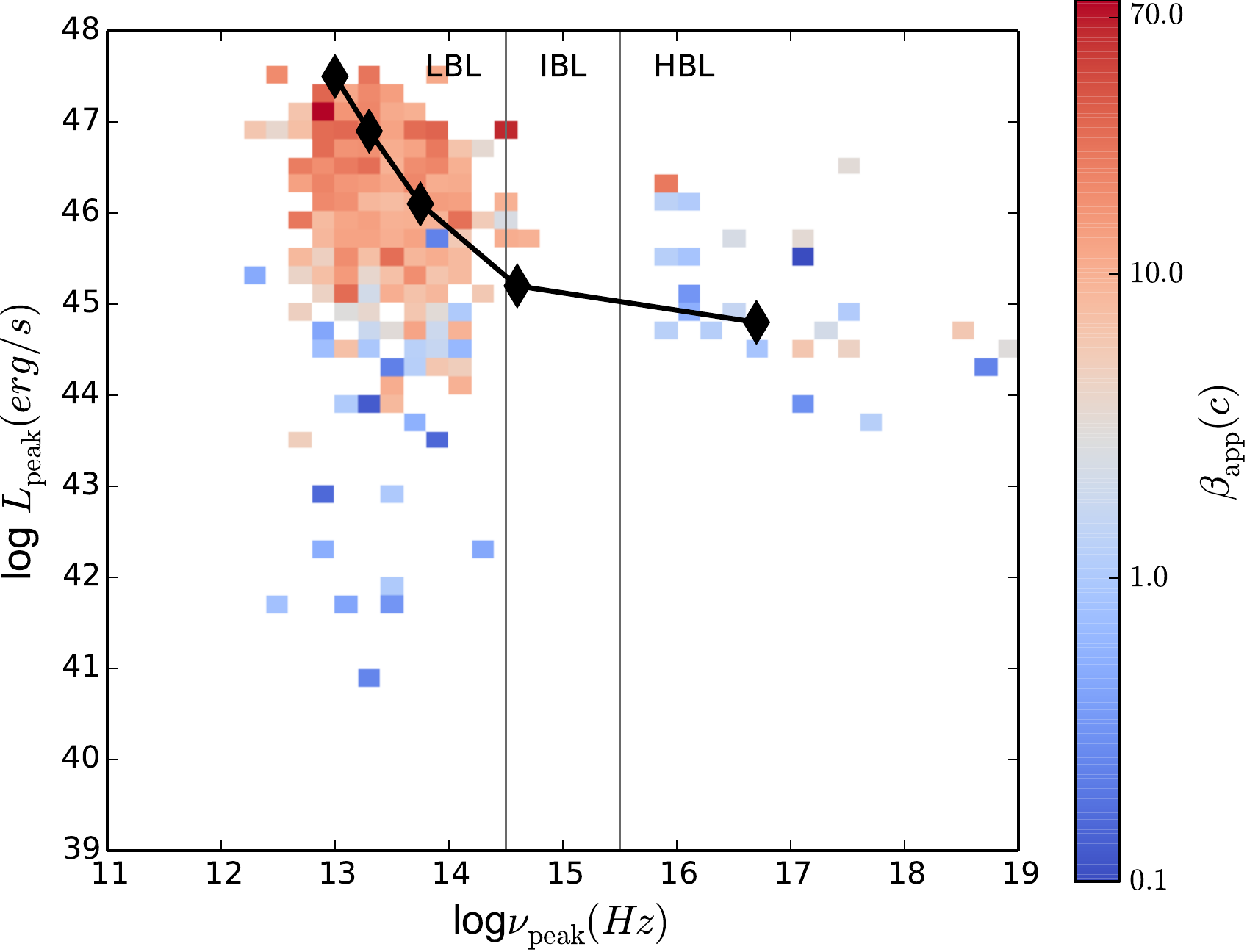}
    \caption{The \vplp ~plane for all sources with measured $\beta_\mathrm{app}$  where we have taken an average over 2D bins in the plane. Red indicates fast speeds, while blue indicates slow, as given by the (log) color scale at right. As expected, there is a noticeable gradient from aligned to misaligned among sources at the left of the plane. We find that the maximum speed also decreases along the original blazar sequence (as indicated by the connected black diamonds).}
    \label{speed_map}
\end{figure}

VLBI studies have also shown that low-power jets have lower observed values of $\beta_\mathrm{app}$ \citep[e.g.,][]{kharb10}, so we also expect to see lower values of  $\beta_\mathrm{app}$ in the type-I (weak jet) region of the \vplp~plane, relative to the type II (strong) zone, as well as the decrease in $\beta_\mathrm{app}$ with increasing orientation angle.  Figure \ref{speed_map} shows a binned average map (made in similar fashion to Figure \ref{fig:orientation_map}) of the \vplp ~plane with a color scale according to average $\beta_\mathrm{app}$, using all sources with measured $\beta_\mathrm{app}$ from the literature (400 total).
%\footnote{The database of VLBI jet speeds has been compiled for another project and will be presented in Keenan et al., in prep.}.  
While there are fewer high-$\nu_\mathrm{peak}$ sources with measured $\beta_\mathrm{app}$, we find that jet speeds are lower in the region of type I blazars and radio galaxies, as expected.

\subsection{LBLs are type~II Jets}
\label{spectraltype}

\begin{figure} 
    \centering
    \includegraphics[width=\linewidth]{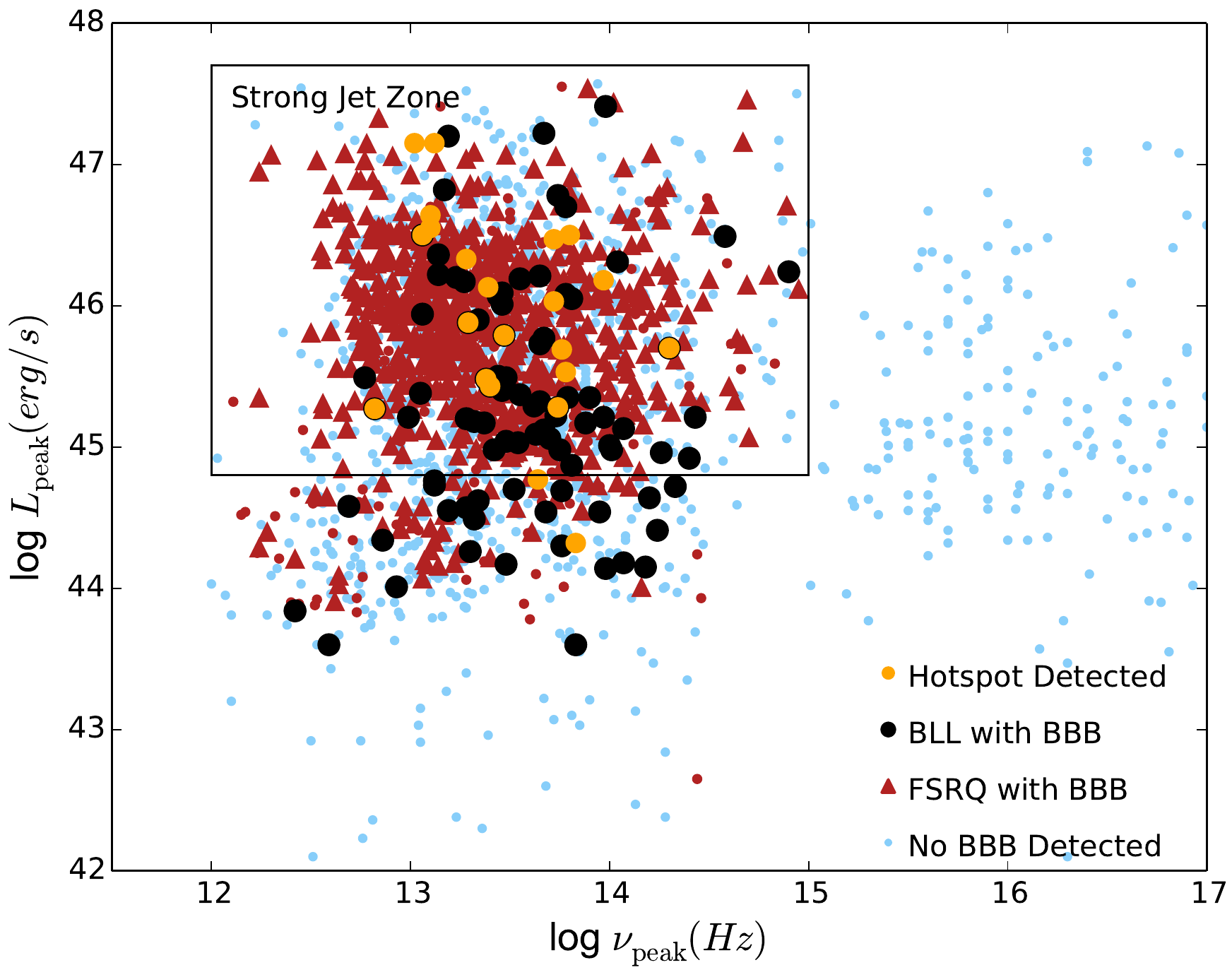}
    \caption{The $\nu_\textrm{peak}-L_\textrm{peak}$ plane for the entire TEX+UEX sample, where we have highlighted sources which have features of type~II jets. Those with a BBB feature are plotted as either red triangles (for FSRQ) or black circles (for BLLs). We also plot as orange circles the location of 22 BLLs from Kharb et al., (2010) noted to have hotspots in high-resolution radio imaging. All the remaining sources (where presence of BBB is unclear) are plotted as small blue dots. All sources with these features, including the BL Lacs, lie in or very near the region we have defined as the strong/type II) zone.}
    \label{fig:bbb}
\end{figure}

Previous authors have suggested that some low-peaking BL Lacs (LBLs, $\nu_\mathrm{peak}\lesssim10^{15}$\,Hz) are actually mis-identified FSRQ, where the broad optical emission lines are present but overwhelmed by the non-thermal jet emission \citep[e.g.][]{markos98,ghis11}. This is supported by the observation that some BL Lac objects (including BL Lac itself) show broad lines when the jet is in a low state \citep[e.g.][]{vermeulen95,corbett96,ghisellini2011,2015MNRAS.449.3517D}, while others have features consistent with type~II jets such as hotspots \citep[][]{kollgaard92,murphy93,kharb10} or the presence of a `big blue bump' accretion disk signature in their broad-band SEDs. In our SED fitting, a BBB component was added to an SED model fit when an obvious blue/UV bump was present in the data (regardless of any other source characteristics). The goal was mainly to improve the synchrotron peak fit by eliminating data from another emission source, but it is also useful to see where jets with a BBB component fall in the \vplp~plane.  In Figure~\ref{fig:bbb} we plot $L_\mathrm{peak}$ vs. $\nu_\mathrm{peak}$ again for the full UEX+TEX sample, and highlight those sources with a clear BBB signature as red triangles (for FSRQ) and black circles (for BL Lacs). We note that the absence of a BBB signature does not mean that the source lacks a strong accretion disk, as many sources do not have sufficient spectral data to be able to detect the BBB. This figure also shows as orange circles the location of 22 BL Lacs in our sample which \cite{kharb10} described as having a hotspot feature in their high-resolution VLA imaging. All other sources are shown with pale blue dots. Essentially \emph{all} the BL Lacs with either a BBB signature or a hotspot occur in the same zone which contains essentially all the broad-lined blazars (e.g., the boxed area which has a maximum $\nu_\mathrm{peak}$ value of $10^{15}$~Hz).\footnote{It is important not to assume that all the blue points in this figure definitely \emph{lack} a BBB -- in many cases the data are not available to make any determination.}

\begin{figure}
    \centering
   \includegraphics[width=\linewidth]{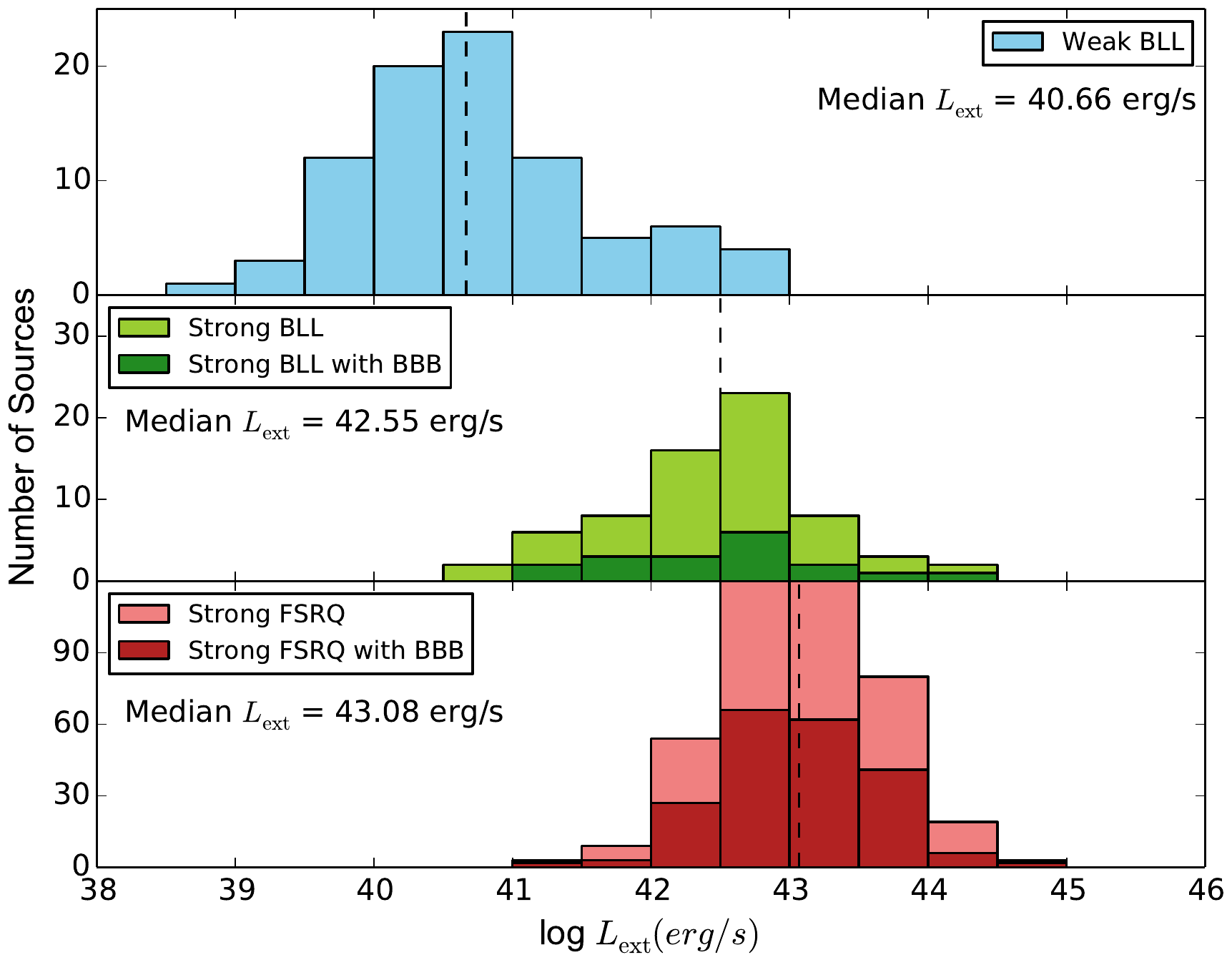}
    \caption{Comparison of the distribution of $L_\mathrm{ext}$ between weak-zone BLLs (defined as log~$\nu_\mathrm{synch}>10^{15}$ Hz) in the upper panel, strong-zone BLLs and FSRQs in the middle and lower panels, respectively. The strong/type~II zone is defined as $\nu_\textrm{peak} < 10^{15}$ Hz and $L_\textrm{peak} > 10^{45.5}$ erg/s.}
    \label{fbl_hist}
\end{figure}

Motivated by these results, we next looked at all the sources located in the strong zone that are classified as BL Lacs, regardless of having a BBB signature or hotspot. We compared these "Type~II BLL" sources (selected with $\nu_\textrm{peak} < 10^{15}$ Hz and $L_\textrm{peak} > 10^{45.5}$ erg/s) to sources in the same zone classified as FSRQ, and to "weak BLL" sources with high synchrotron peaks ($\nu_\textrm{peak} > 10^{15}$ Hz).  The distributions of $L_\textrm{ext}$ for these samples is shown in Figure \ref{fbl_hist}, where we also show the sources with a BBB as the darker shaded part of the histogram. The dotted lines represent the average value of $L_\mathrm{ext}$ for each sample. 
It is clear that the mean and distribution of $L_\mathrm{ext}$ for strong-zone BLLs is far more similar to FSRQ than to the high-$\nu_\mathrm{peak}$, weak-zone BLLs. Given these observations, as well as the fact that many LBLs have extended radio powers well above what are observed for FR I radio galaxies \citep[as has been noted previously, e.g.][]{rector01}, it seems that LBLs should be classified as type~II jets along with FSRQ.

%Our results here confirm that the zeroth order unification scheme aligning BL Lacs with FR Is and FSRQ with FR IIs is in need of amendment, and that LBLs are better grouped with FSRQs as strong-jet sources. 
Interestingly, putting LBLs in the type II class with FSRQ and FR~II may solve a longstanding issue of conflicting measurements of cosmological evolution in BL Lac samples.  FR I jets samples are typically found to exhibit negative evolution \citep{maraschi94,perlman04}, while FR IIs and FSRQs exhibit positive evolution, similar to radio-quiet quasars \citep{giommi2009,mao2017}. BL Lac samples have had mixed results, but those which have reported slightly positive or neutral evolution include LBLs \citep[e.g.,][]{rector01,caccianiga02}, while the strongest negative evolution has been observed for HBL samples \citep{bade98,giommi99,ajello14}.  If many LBLs are actually positively-evolving type~II jets, then it may be the case that the inclusion of these sources in BL Lac samples has masked the negative evolution of the high-peaking BL Lacs which are the true match sample for FR I radio galaxies.

\subsection{Accretion Mode in type I and II Jets}
\label{accretion_divide}

\begin{figure}
    \centering
    \includegraphics[width=\linewidth]{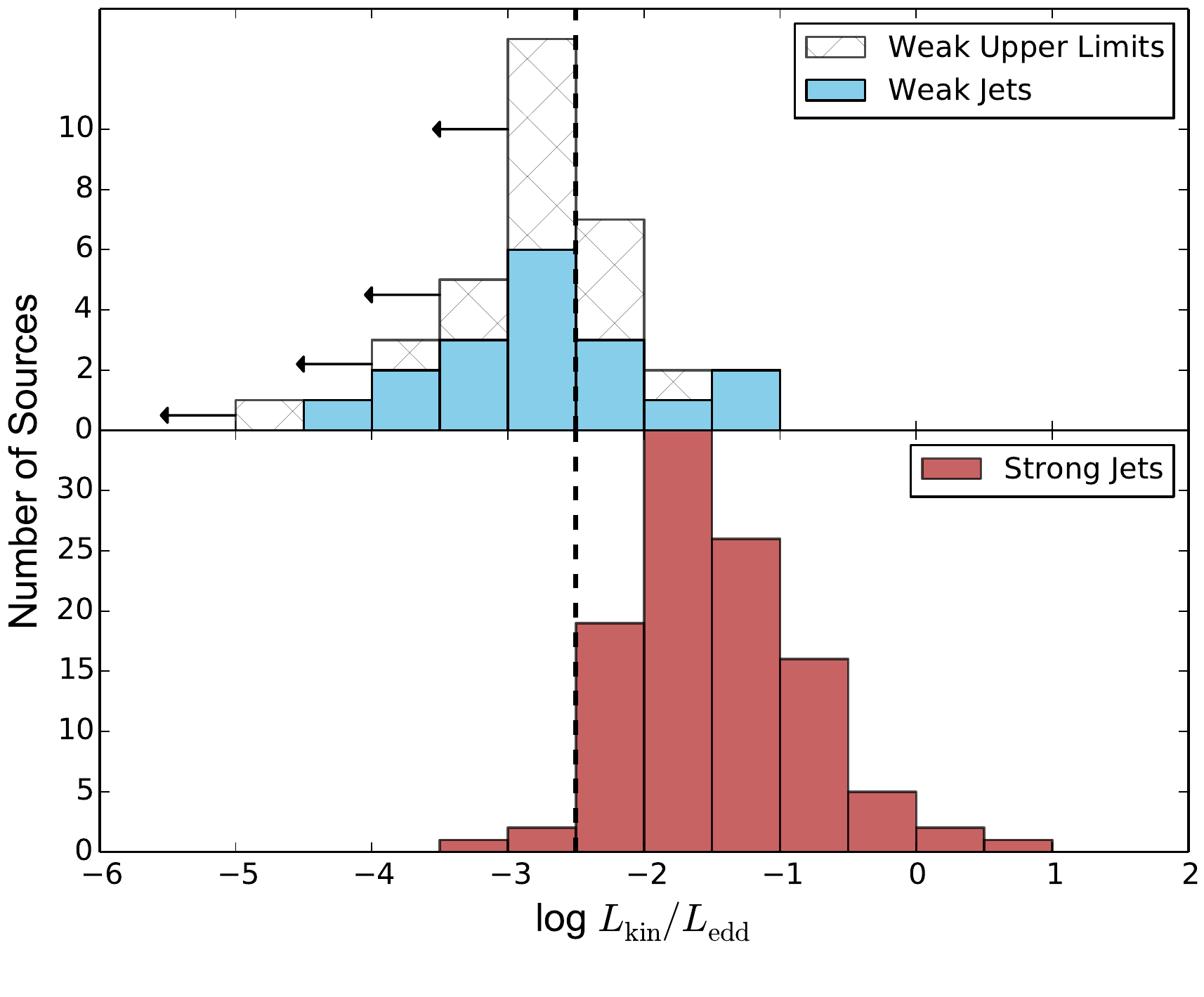}
    \caption{Histogram showing a comparison of log $L_\mathrm{kin}/L_\mathrm{edd}$ (the Eddington ratio) between type I/II jets (selected based on position in the $\nu_\mathrm{peak}-L_\mathrm{peak}$ plane). The type~I sample includes sources with upper limits on $L_\mathrm{ext}$ due to the small numbers of sources. A dashed vertical line is drawn at log $L_\mathrm{kin}/L_\mathrm{edd}=-2.5$ where we see a divide in the population.}
    \label{acc_hist}
\end{figure}

It is clear that jet power and orientation shape the distribution of sources in the $\nu_\textrm{peak}$--$L_\textrm{peak}$ plane, but the very large range of overlap in jet power between type I and II jets suggests that it alone does not predict the type of jet produced.  
It has long been hypothesized that accretion mode could drive a dichotomy in jetted AGN \citep[e.g.][]{ghisellini01,2003ApJ...593..667M,2004MNRAS.351..733M,2006MNRAS.370.1893H,2009MNRAS.396L.105G,meyer2011,antonucci}.  
Here we use the Eddington ratio, log $L_\textrm{kin}/L_\textrm{edd}$, as a proxy for the accretion efficiency, where $L_\textrm{kin}$ is calculated from $L_\mathrm{ext}$ using the scaling presented in \cite{cavagnolo10}: $$\log L_\mathrm{kin} = 0.64 (\log L_\mathrm{ext} - 40)
+ 43.54 \mathrm{ (ergs/s).}$$ As before, we identified sources as type~II/strong ($L_\textrm{peak} > 10^{45.5}$ erg/s and $ \nu_\textrm{peak} < 10^{15}$ Hz) and type~I/weak ($\nu_\textrm{peak} > 10^{15}$ Hz).  Neither of these identifications take into account the original classification (as BLL, FSRQ, etc) or $L_\textrm{ext}$ value, simply the location of the source in the $\nu_\textrm{peak}$-$L_\textrm{peak}$ plane.  Figure \ref{acc_hist} shows a comparison of log $L_\textrm{kin}/L_\textrm{edd}$ between these samples. Because there are few weak-zone sources with measured black hole mass, we have included those with upper limits on $L_\textrm{ext}$ in this sample, as noted by the gray-crossed portion of the upper histogram.  In agreement with previous work, we see a divide in Eddington ratio at $L_\textrm{kin}/L_\textrm{edd}\approx-2.5$.

We found no significant difference in the distributions of $M_\textrm{BH}$ between our samples of type I/II jets. Both populations peak at $M_\textrm{BH} \approx 10^{8.5-9} M_\odot$ (histograms are shown in the online supplemental Figure~\ref{supp:mbh_hist}).  This agrees with previous work showing no connection between black hole mass and accretion efficiency \citep[e.g.][]{2002ApJ...580...96O,2004MNRAS.351..733M}. The similarity of black hole masses can explain both the large range in overlap in jet powers for type I and II jets, and the upper bound on jet power in type I jets. For the latter, we can predict this upper bound by simply calculating the jet power for a ``maximum inefficient'' Eddington fraction of 5\% for a $10^{9.5}\,M_\odot$ black hole. Using the \cite{cavagnolo10} scaling, this corresponds to a predicted maximum value of log~$L_\mathrm{ext}=10^{42.7}$ erg/s for weak/type I sources which matches near-exactly our observations. We expect, then, that the high-$\nu_\mathrm{peak}$, high-$L_\mathrm{peak}$ sources in the blazar sequence plane, which have the highest jet powers of all type I jets, will predominantly be hosted by the most massive black holes and be accreting at the `maximum' for inefficient accretion, of about $0.5-1$\% Eddington. Some signs of this do appear in the plot of $L_\mathrm{ext}$ versus $M_\mathrm{BH}$ for weak jets (see online Supplemental Figure~\ref{supp:lextmbh_weak}), but the small number of measurements compared to upper limits gives us low statistical power to detect a correlation.  

A similar argument explains why type II jets may exist with very low jet powers -- these may simply be hosted by smaller black holes.  We find  a marginally significant correlation between black hole mass and $L_\mathrm{ext}$ (Kendall's $\tau$=0.13 with a p-value of 0.04, see online Supplemental Figure~\ref{supp:lextmbh_strong}). Given the very large error on both of these measurements and the fact that efficiency may vary by a factor of 10 or more even within type II jets, it is not surprising that the correlation is difficult to discern, assuming it is real. Follow-up studies focusing on sources with very well-measured jet power and black-hole mass, and with an independent means of estimating the accretion rate, may be able to better test for the expected correlations in the two jet classes.

\section{Conclusions}
\label{sec:conclusions}
In this work, we have compiled a very large sample of more than 2000 SEDs of the non-thermal jet emission in radio-loud AGN, from highly aligned blazar jets to the more misaligned jets hosted by radio galaxies. An important component of this work has been the careful measurement of the extended (isotropic) radio luminosity of the jet for over 1000 sources, which we use to estimate the jet power. With this sample, we have revisited the synchrotron peak luminosity - peak frequency (\vplp) plane in which \cite{fossati98} famously discovered an anti-correlation more than 20 years ago. With 15 times as many sources as in the original blazar sequence, it is now clear that the blazar sequence did likely arise partly from selection effects, as has long been argued. We find that there is no `forbidden zone' at high-$\nu_\mathrm{peak}$ and high $L_\mathrm{peak}$, but rather that this area of the plane is populated by type I ("weak" or inefficiently accreting) BL Lacs. These high-$\nu_\mathrm{peak}$ and high $L_\mathrm{peak}$ sources may be rare cases of very massive black holes ($M_\mathrm{BH}\approx 10^{9.5} M_\odot$) accreting inefficiently but close to the `switch' at about 0.5\% of Eddington, consequently leading to moderately high jet power, which is correlated with synchrotron peak power.

One of the main attractions of the blazar sequence paradigm was the idea that jets might be monoparametric, with their spectral type and broad SED characteristics predicted by a single parameter, the jet power. However we have shown that outside of the extreme high-power tail of the population, this is not the case. Indeed, a source with low to moderately high power (the bulk of the population) may by type I or type II, and may have a synchrotron peak anywhere between $10^{12}$ and $10^{17} Hz$. The power alone does not seem to specify spectral type in any sense for most jets.

With the blazar sequence set aside, we have re-examined the \vplp~ plane to see what normalities and trends still remain.  Based on our observations, we suggest two classes of jets, type I and II (i.e., `weak' and `strong' jets). Type II jets we assume are those hosted by systems with efficient (or `quasar mode') accretion, while type I jets are associated with inefficient/highly sub-Eddington accretion, as has been argued by \cite{2009MNRAS.396L.105G} based on observations of gamma-ray blazars. We find that for the relatively aligned jets in these two classes, their synchrotron spectra differ greatly, and the two populations can be neatly separated in the \vplp~plane, with the strong/type II zone defined as $L_\mathrm{peak} > 10^{45}$ erg/s and $\nu_\mathrm{peak}< 10^{15}$ Hz, and the weak/type I zone as $\nu_\mathrm{peak} > 10^{15}$ Hz.  

\vspace{8pt}
\noindent
With this classification system in mind, our key observations and conclusions are:

{
\addtolength{\leftskip}{5mm}

\vspace{3pt}
\noindent
\emph{(i)} Type II jets (as represented by the subset that are FSRQ/FR II) have a clear upper bound on their synchrotron peak frequency of $\nu_\mathrm{peak}\approx10^{15}$\,Hz, with 99\% of the sample below this value. Since these jets have broad-line-emitting gas in the nuclear region, this may be due to the synchrotron cooling effect of a dense external photon field, as first suggested by \cite{ghisellini98}. 
    
\vspace{3pt}
\noindent
\emph{(ii)} Type I jets have no such bound on $\nu_\mathrm{peak}$, and can have values ranging up to $10^{19}$\,Hz at the most extreme. The wide range of synchrotron peak frequency and luminosity among type I jets can in part be explained through the wide range in jet power as well as velocity gradients in the jet, which is likely influenced by their environment.

\vspace{3pt}
\noindent
\emph{(iii)} Low-synchrotron-peak BL Lacs (LBLs) are mostly type II, and jet studies that contrast broad-lined FSRQs with BL Lacs should group LBLs with FSRQ as type II jets. 

\vspace{3pt}
\noindent
\emph{(iv)} Type II jets exist over four decades in jet power, down to the very lowest values observed for the full jet sample ($L_\mathrm{ext}<10^{41}$erg\,s$^{-1}$). We propose that low-power type~II jets are produced by smaller and/or minimally accreting, but still "efficient" black holes. Additional observations are needed to confirm this.

\vspace{3pt}
\noindent
\emph{(v)} Weak jets have an upper bound jet power of approximately $10^{43}$erg\,s$^{-1}$ which corresponds to the expected bound for a "maximal" inefficient source with an Eddington fraction of 0.5-1\% onto the most massive black holes ($M_\mathrm{BH}$ = $10^{9.5}-10^{10}\, M_\odot$).

}

\vspace{8pt}

There remain many directions for future study. A more precise understanding of velocity gradients in weak jets and the role of environment to make sense of the exact location of type I jets in the \vplp~plane (where these jets spread over at least 4 orders of magnitude in each quantity) requires further study. 
%In a future paper, we will also examine the properties of the high-energy Compton emission from these jets, which we have not yet explored. Finally, 
While we have included as many misaligned jets (i.e., radio galaxies) as possible, those with well-sampled jet SEDs remain rare, as is also the case for low-power jets like those increasingly found in LLAGN (some of which are formally radio-quiet). A more accurate characterization of the jet SED in radio galaxies may reveal differences in their synchrotron properties which were not readily apparent in this work. Such a study should also explicitly take into account the excitation index. Indeed, the importance of jet morphology in radio galaxies in our proposed unification scheme is not entirely clear. Recent work showing that the HERG and LERG classes both exhibit FR I and FR II morphologies suggests that environment may play a larger role than previously thought. Because very high jet power (e.g. $L_\mathrm{ext} > 10^{43}$ erg/s) always selects a type II jet, it is probably safe to assume that the most powerful FR IIs are indeed the misaligned counterparts of type II blazars.  At moderate powers, however, it is likely better to make the distinction between excitation type.

\section*{Acknowledgements}

The authors thank the anonymous referee for extensive comments that improved the paper.

MEK acknowledges a GAANN Fellowship from the Department of Education (P200A150003). 

This research has made use of the SIMBAD database, as well as the VizieR catalog access tool managed at CDS, Strasbourg, France.   

This research has also made use of the NASA/IPAC Extragalactic Database (NED), which is operated by the Jet Propulsion Laboratory, California Institute of Technology, under contract with the National Aeronautics and Space Administration.  

This paper makes use of the following ALMA data: ADS/JAO.ALMA\#2015.1.00932.S, ADS/JAO.ALMA\#2016.1.01481.S, and ADS/JAO.ALMA\#2017.1.01572.S. ALMA is a partnership of ESO (representing its member states), NSF (USA) and NINS (Japan), together with NRC (Canada), MOST and ASIAA (Taiwan), and KASI (Republic of Korea), in cooperation with the Republic of Chile. The Joint ALMA Observatory is operated by ESO, AUI/NRAO and NAOJ.  

The National Radio Astronomy Observatory is a facility of the National Science Foundation operated under cooperative agreement by Associated Universities, Inc.  

This research has made use of data from the MOJAVE database that is maintained by the MOJAVE team \citep{lister2018}.

This research made use of Astropy,\footnote{http://www.astropy.org} a community-developed core Python package for Astronomy \citep{astropy:2013, astropy:2018}. 

\vspace{12pt}
\noindent
\textbf{Data Availability Statement.} The data underlying this article are available in the article and in its online supplementary material.
%%%%%%%%%%%%%%%%%%%%%%%%%%%%%%%%%%%%%%%%%%%%%%%%%%



\bibliographystyle{mnras}
\bibliography{reference} % if your bibtex file is called example.bib

%\begin{thebibliography}{99}
%\end{thebibliography}

%%%%%%%%%%%%%%%%%%%%%%%%%%%%%%%%%%%%%%%%%%%%%%%%%%

%%%%%%%%%%%%%%%%% APPENDICES %%%%%%%%%%%%%%%%%%%%%

\clearpage
\appendix
\section{}
\subsection{Comparing the Poorly Sampled Jets to TEX/UEX Samples}

Given that our main analysis is based on about 1/3 of our starting sample of nearly 7000 potential sources, it is important to compare the properties of the jets with well-constrained SEDs (the TEX and UEX samples) to the 4732 sources remaining in the starting sample. About half these sources were discarded without SED processing for the following reasons: 129 sources were stellar/non-AGN sources according to SIMBAD (e.g., white dwarfs or X-ray binaries -- these can contaminate blazar candidate catalogs based on multi-band selection), 1771 had fewer than 3 radio ($<10^{13}$\,Hz) data points, and 1312 had fewer than 3 data points at higher frequencies. The remaining 1520 sources met the minimum data requirements to attempt to fit/assess the SED but were ultimately rejected.  In general, these sources did not make it into either the UEX or TEX sample due to insufficient multi-wavelength spectral coverage to verify its identity as a jetted AGN and/or constrain the synchrotron peak. In many cases the optical/UV data points are clustered in a narrow range or consist of a single data point; given that the optical/UV is the most likely part of the spectrum to show spectral components from other parts of the AGN (e.g., host, accretion disk), in these cases it is difficult to verify that the optical fluxes are dominated non-thermal emission from the jet, much less use the limited data to constrain the synchrotron peak. Indeed, classifiers noted at least 400 of examined-but-rejected sources are clearly misaligned blazars/RG with SEDs dominated by their `AGN' components (e.g., thermal dust/torus, stellar continuum from the host galaxy, accretion disk in the UV). Some misaligned jetted AGN with careful isolation and fitting of the nuclear jet emission have been retained in the main samples as noted in Section~\ref{sec:sedfit}, but most misaligned jets require further careful study and more complex SED fitting to properly isolate AGN and jet components. 

Putting aside the 129 non-AGN sources and 7 of unknown type leaves 4528 sources we refer to as the `rejected' or `dropped' sources which we will further examine. Of these 71\%  have a clear identification as an AGN of some type, 25\% are identified only by a single band (e.g., `gamma-ray source', `radio source'), and 4\% are identified by SIMBAD (possibly incorrectly) as non-active galaxies. Of the AGN, about 2/3 appear to have a blazar type (1414 primary SIMBAD type of `BLL' and 759 of `QSO').  For the rejected sample, a large fraction lacked X-ray observations (76\%) or had wide gaps in coverage between radio ($\nu\sim10^9$\,Hz) and the far-IR ($\nu\sim10^{14}$\,Hz, 50\%), and nearly half (45\%) lack redshifts.

\begin{figure}
    \centering
    \includegraphics[width=3in]{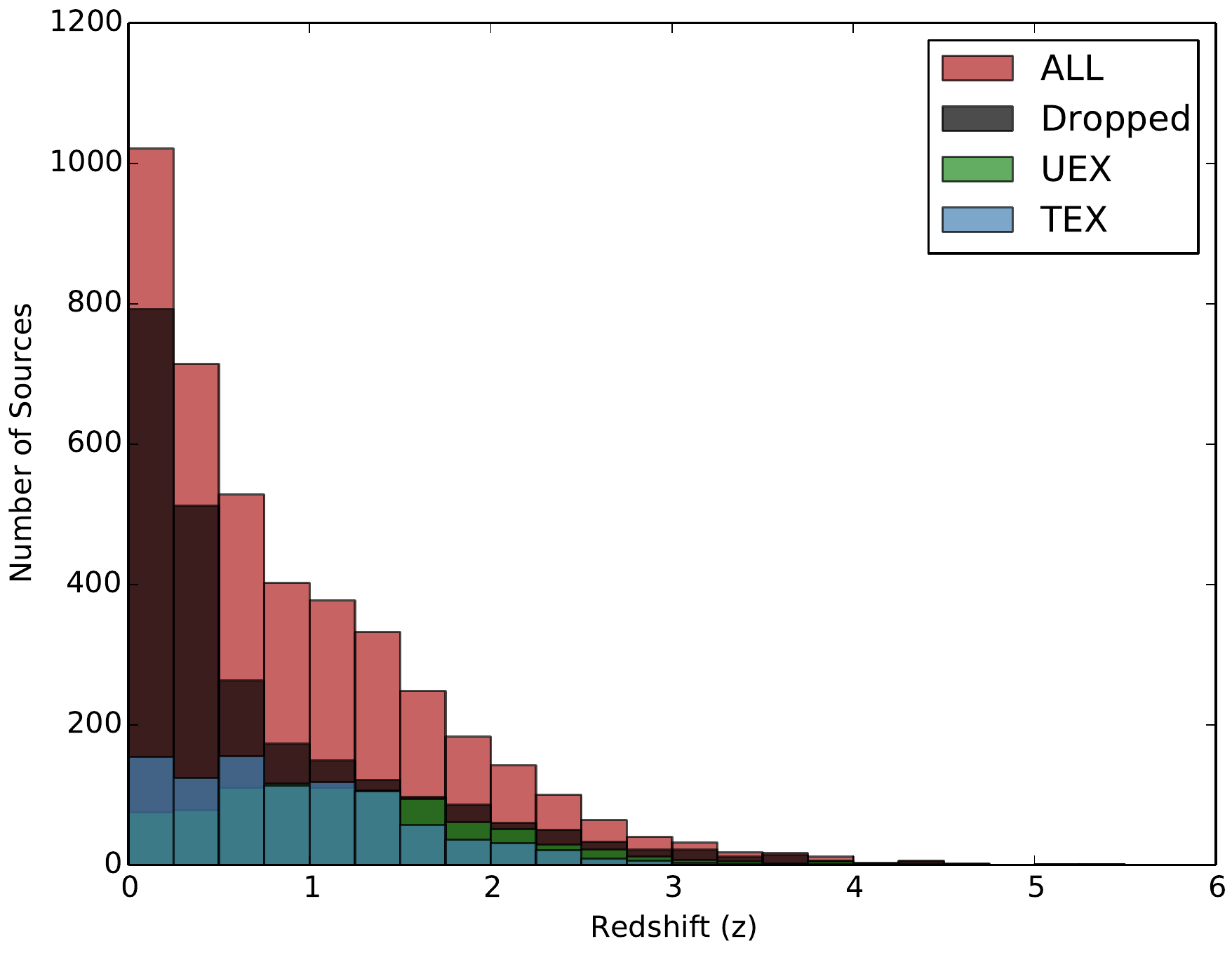}
    \caption{Histograms showing the redshift distributions of the TEX, UEX, discarded and total samples. We find that while the TEX and UEX have very similar distributions, there is a surplus of sources with low redshift which were discarded due to insufficient spectral data.}
    \label{appA:redshift}
\end{figure}

\begin{figure}
    \centering
    \includegraphics[width=3in]{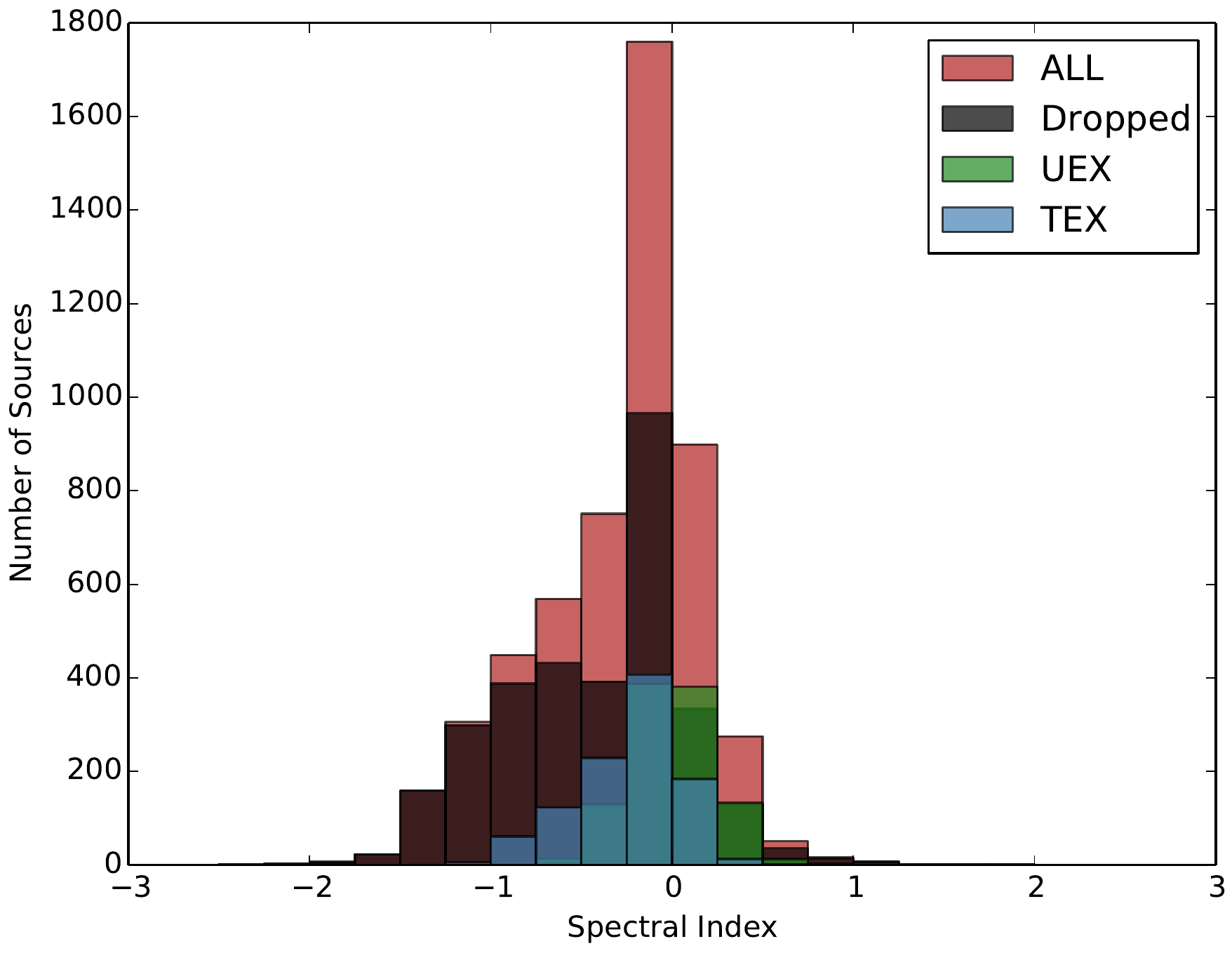}
    \caption{Histograms showing the spectral index distributions for the TEX, UEX, discarded and total samples. We find that the TEX and UEX have similar distributions, with TEX having a slight preference for steep spectrum sources. There is a surplus of sources with very steep spectra that were discarded due to insufficient SED sampling.}
    \label{appA:spectral_ind}
\end{figure}

In Figure \ref{appA:redshift}, we compare the redshift distributions of the TEX, UEX, and discarded sources.  We find that the TEX and UEX have similar redshift distributions, while the rejected sample has a surplus of low-redshift sources.\footnote{Note that 103/1045, 169/1079 and 2074/4532 of the TEX, UEX, and rejected samples have unknown redshift, respectively} Figure \ref{appA:spectral_ind} compares the radio spectral indices of the samples.  The spectral indices were calculated via a simple least squares fitting for sources with at least 2 fluxes below $10^{13}$ Hz; here we adopt the convention $F(\nu) \propto \nu^{a}$. We find that the TEX and UEX have similar distributions, with a slight preference for steep sources in the TEX and flat sources in the UEX samples. This is expected if at least some fraction of the UEX sources are highly-aligned blazars that are jet-dominated even down to low radio frequencies, which prevents us from using spectral decomposition to measure the extended radio power.  There is a surplus of sources with very steep spectra in the discarded sample, which suggests this sample contains more misaligned jets. 

\begin{figure}
    \centering
    \includegraphics[width=3in]{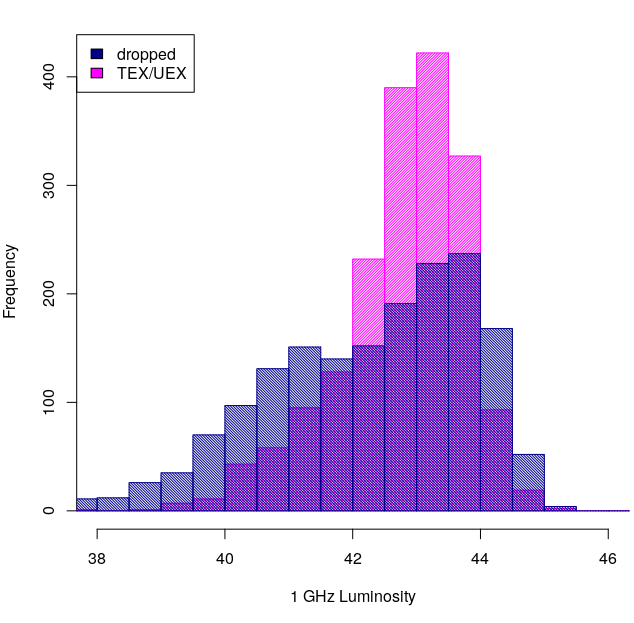}
    \caption{Histograms showing the radio luminosity distributions for the TEX/UEX (magenta) and dropped (blue) samples (average 1 GHz value taken from NED, where redshifts are available). While the dropped sample includes a larger proportion of low-power sources, there is also a small population of higher-power sources, predominantly high-redshift quasars or relatively nearby powerful radio galaxies. }
    \label{appA:lum}
\end{figure}

Together we interpret Figures~\ref{appA:redshift} and \ref{appA:spectral_ind} to imply that the discarded sources have a higher share of more local ($z<1$) low-power, misaligned sources, which is confirmed in Figure~\ref{appA:lum}. Given that even misaligned jetted sources have higher fluxes in the radio than at higher frequencies, it is not surprising that this population is more likely to be dropped due to the lack of matching deep optical and X-ray surveys. The radio data available for the rejected sample varies widely in coverage and quality, but we can look at the 1 GHz radio luminosity for sources with redshifts in comparison to our TEX/UEX sample in Figure~\ref{appA:lum}.  The dropped sample also includes a number of apparently high-luminosity sources (log$_\mathrm{10}$ $L_\mathrm{1GHz}>44$). These are almost entirely high redshift QSOs (143 sources, $<z>$=2.4) with no X-ray flux (75\%) or relatively high-redshift radio galaxies (45 sources, $<z>$=1.7).  

We now ask what might be the effect of these missing sources on our general observations and conclusions from the well-characterized sample. We consider these in turn:

1) The existence of high-$\nu_\mathrm{peak}$, high-$L_\mathrm{peak}$ sources, as well as the population of low-power type II (strong) jets, which are counter to the idea of a monoparametric blazar sequence. Since our observations already seem to indicate that a monoparametric ``sequence'' in the $\nu_\mathrm{peak}-L_\mathrm{peak}$ plane is not possible, it is not clear how this population could overturn this observation.  Even if every discarded source followed something like the blazar sequence, a significant population of outliers is already established in the UEX/TEX samples (and in other works in the literature). If the discarded sources consist of more outliers (as seems likely, given recent observational work which has discovered a population of local, low-power FR II jets), then our argument against a monoparametric blazar population is strengthened.

2) For our well-characterized sample, we find that essentially zero sources with apparent type II accretion activity (e.g., with broad lines or a BBB) have $\nu_\mathrm{peak}$ values above $10^{15}$\,Hz.  
%It is possible, though not likely, that broad-lined sources with high nu_peak values uniquely exist in the rejected sample. One reason that this appears unlikely is that the same observation (no high-efficiency-accretion signs above 10^15) was made in Fossati '98 with a little over 100 sources, again in Meyer 2011 with over 700 sources, and finally here with over 2000 sources. It does not seem credible that despite increasing our samples of well-characterized SEDs, these sources exist but are eliminated due to selection effects, 
Since no selection is made on the existence of broad lines or a BBB to be included in our well-characterized sample, it is not likely that these sources exist \emph{uniquely} in the rejected sample. Indeed, there is no reason why we would not see the BBB or broad lines in at least some of our sources with $\nu_\mathrm{peak}>10^{15}$\,Hz if they exhibited either.

%In these sources, the extended emission would dominate the low frequency spectrum and the core emission would be boosted in a direction nearly perpendicular to the line of sight, thus making it extremely difficult to measure the core spectrum necessary for inclusion in the TEX or UEX samples. We would expect many of these sources to be in the lower left corner of the \vplp ~plane. \textbf{radio flux vs redshift plot}

3) The "upper limit" on jet power in type I (weak) jets. We have observed that the highest jet powers reached for sources in the "weak jet" zone ($\nu_\mathrm{peak}>10^{15}$\,Hz) naturally corresponds to the expected "maximal but still inefficient" accretion rate onto a maximally-sized ($\sim10^9$ $M_\odot$) black hole, and that these  high-$\nu_\mathrm{peak}$, high-$L_\mathrm{peak}$ sources do appear to have high $M_\odot$. Could there be a population of sources with similar synchrotron properties (i.e., also in the weak jet zone) but even higher jet powers? There is no obvious reason why these would be entirely absent from our sample of over 2000, if similar but lower-power sources are already included. It seems likely that such powerful sources would minimally appear in the UEX sample, but the upper limits on the extended power for that sample precludes this (as discussed in the main text, see also online Figure~B3).  
%Further, we see from Figure~A3 that the rejected population are predominantly low-luminosity sources.

\subsection{Sources Lacking Redshifts in the TEX/UEX Samples}

\begin{figure}
 \includegraphics[width=3in]{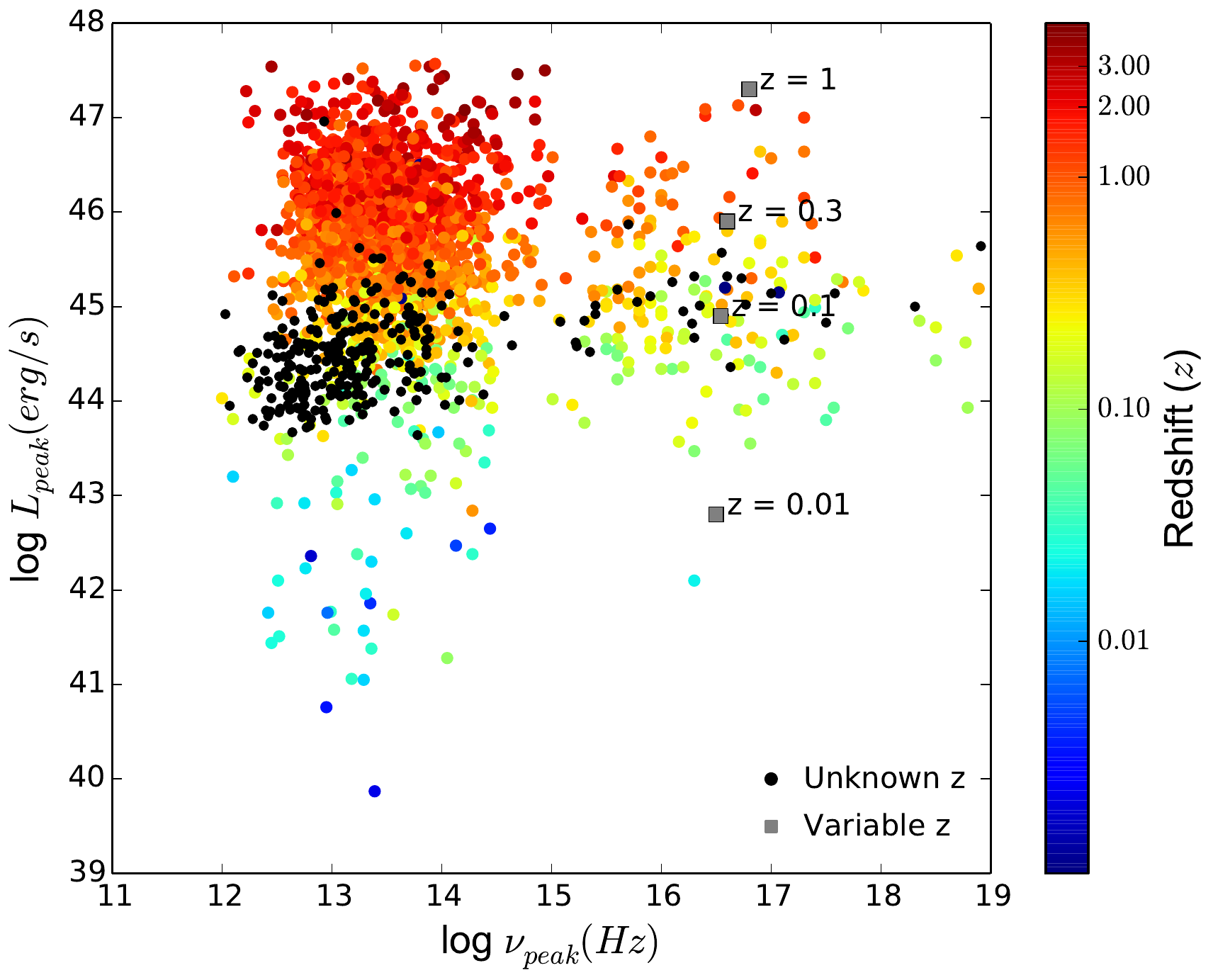}
 \caption{The \vplp~plane for both the TEX and UEX samples with a color scale corresponding to redshift ($z$).  All sources with no known redshift are shown as black circles, where we have adopted $z=0.3$.  The gray squares represent a theoretical source with a peak synchrotron flux of 0.05 Jy and peak synchrotron frequency of $10^{16.5}$ Hz, where $L_\mathrm{peak}$ and $\nu_\mathrm{peak}$ are calculated for different values of the redshift. As shown, the assumed redshift makes very little difference to the peak frequency value.}
 \label{fig:supp_redshift}
\end{figure}

In Figure~\ref{fig:supp_redshift} we show the full TEX+UEX sample in the \vplp~plane, with a logarithmic color scale on redshift. The sources which have good SED sampling have been placed on the plot as black dots, assuming a redshift $z=0.3$. We further demonstrate the strong dependence of peak luminosity on the chosen redshift by a sequence of gray square points, as marked. Note that the peak frequency is only slightly affected. Interestingly, the bulk of the unknown-$z$ sources have low synchrotron peak frequencies. 

\section{Supplemental Online Figures}

\begin{table*}
    \centering
    \caption {Properties of the UEX sample of jets with undetected extended radio power} 
    \begin{tabular}{c c c c c c c c c c c}
        Source Name & RA & DEC & Redshift & Type & $L_\textrm{ext}$ & $L_\textrm{peak}$ & $\nu_\textrm{peak}$ & $M_\textrm{BH}$ & Ref. & Sample ID \\
        -- & J2000 & J2000 & -- & -- & erg/s& erg/s & Hz & $M_\odot$ & -- & --\\
        (1) & (2) & (3) & (4) & (5) & (6) & (7) & (8) & (9) & (10) & (11)  \\
        \hline
                       QSO B0414+009 & 04 16 52.49 & +01 05 23.90 & 0.2870 &        BLL & 41.12 & 45.67 & 16.90 &  8.4 &          r &        e,f,n,r,u	\\
                 1Jy 0454-81 & 04 50 05.44 & -81 01 02.23 & 0.4440 &       FSRQ & 41.15 & 45.68 & 12.79 &  8.1 &          p &                h	\\
                TXS 0706+592 & 07 10 30.07 & +59 08 20.37 & 0.1250 &        BLL & 40.48 & 44.85 & 18.35 &  8.4 &    a,p,r,q &        e,f,l,n,r	\\
                 B3 0729+391 & 07 33 20.84 & +39 05 05.15 & 0.6637 &       FSRQ & 41.68 & 45.38 & 13.47 &  8.3 &          c &                q	\\
               QSO B0737+744 & 07 44 05.37 & +74 33 58.26 & 0.3140 &        BLL & 40.63 & 45.36 & 17.00 &  8.5 &          r &    e,f,l,n,p,s,u	\\

        \hline \\
    \end{tabular}
    \begin{tablenotes}
    \item Table \ref{supp:uex_tab1} is published in its entirety in the machine-readable format.  A portion is shown here for guidance regarding its form and content. All measurements in this table other than redshift are $\log_{10}$ values. The letters in column (10) denote the following references for $M_{BH}$ measurements: 
    (a) \cite{2003ApJ...583..134B} 
    (b) \cite{2015PASP..127...67B}
    (c) \cite{2016yCat..74543864B}
    (d) \cite{2009MNRAS.397.1713C}
    (e) \cite{2017ApJ...850...74K}
    (f) \cite{2017ApJS..228....9K}
    (g) \cite{2006ApJ...642..711L}
    (h) \cite{2006ApJ...637..669L}
    (i) \cite{2010MNRAS.407.2399M}
    (j) \cite{2005MNRAS.361..919P}
    (k) \cite{2011MNRAS.413..805P}
    (l) \cite{2017A&A...598A..51R}
    (m) \cite{2018A&A...614A.120S}
    (n) \cite{2016ApJ...831..134V}
    (o) \cite{2004ApJ...615L...9W}
    (p) \cite{2002ApJ...579..530W}
    (q) \cite{2005ApJ...631..762W}
    (r) \cite{2002A&A...389..742W}
    (s) \cite{2004A&A...424..793W}
    (t) \cite{2005AJ....130.2506X}
    
    \end{tablenotes}
    \label{supp:uex_tab1}
\end{table*}

\begin{figure*}
 \includegraphics[width=6in]{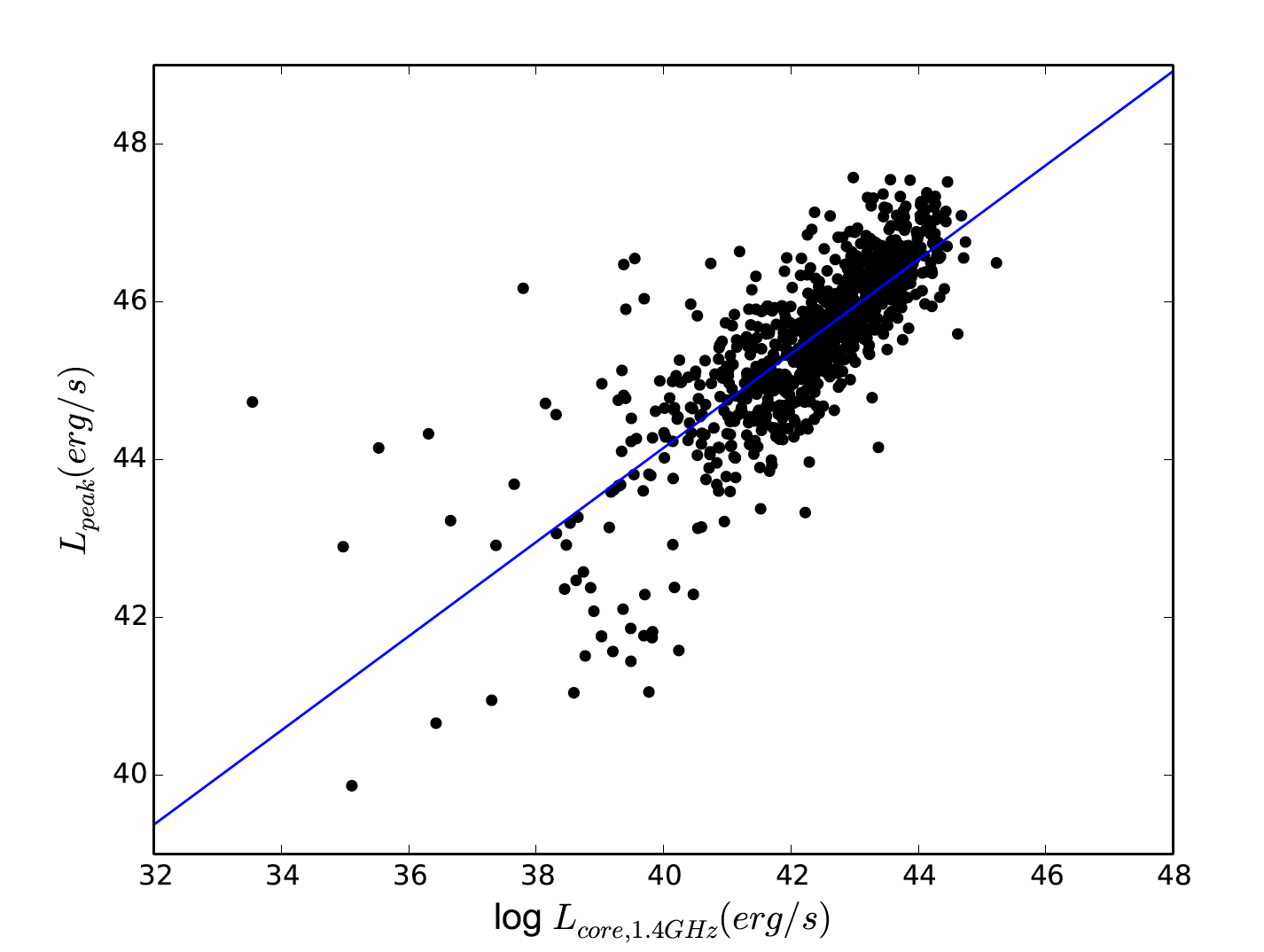}
   \caption{A plot of $L_\mathrm{1.4,core}$ versus $L_\mathrm{peak}$ for the jets in our sample with well-sampled SEDs. The fit relation shown has been used to estimate $L_\mathrm{peak}$ for some radio galaxies with insufficient SED coverage for the parametric SED model fitting.}
   \label{fig:supp_lcorelpeak}
\end{figure*}

\begin{figure*}
    \includegraphics[width=6in]{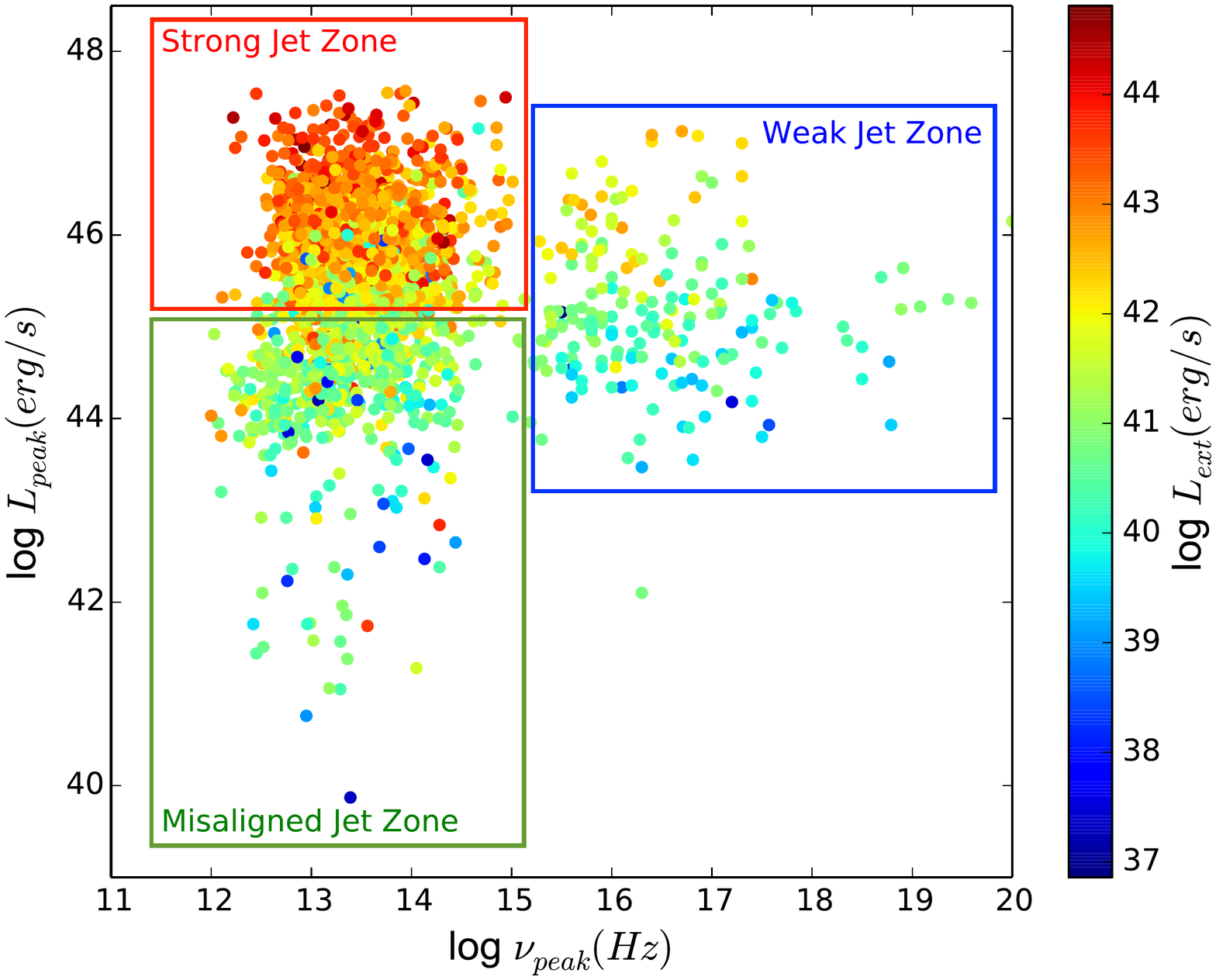}
    \caption{A plot of synchrotron peak luminosity versus peak frequency for the 1055 jets in the TEX sample. Color scale is on log $L_\mathrm{ext}$ as shown at right. Here we show the `zones' for strong and weak jets described in the text.}
    \label{fig:supp_boxes}
\end{figure*}

%\begin{figure*}
%    \centering
%    \includegraphics[width=\linewidth]{envelope_uex.pdf}
%    \caption{The UEX sample in the \vplp ~plane with the color scale representing upper limit of $L_\mathrm{ext}$. We find that none of the high-$L_\mathrm{peak}$, high-$\nu_\mathrm{peak}$ sources have values in the two highest energy bin of Figure \ref{4panel} ($L_\mathrm{ext} > 42$).}
%    \label{fig:supp_uex_uls}
%\end{figure*}

\begin{figure*}
    \centering
    \includegraphics[width=\linewidth]{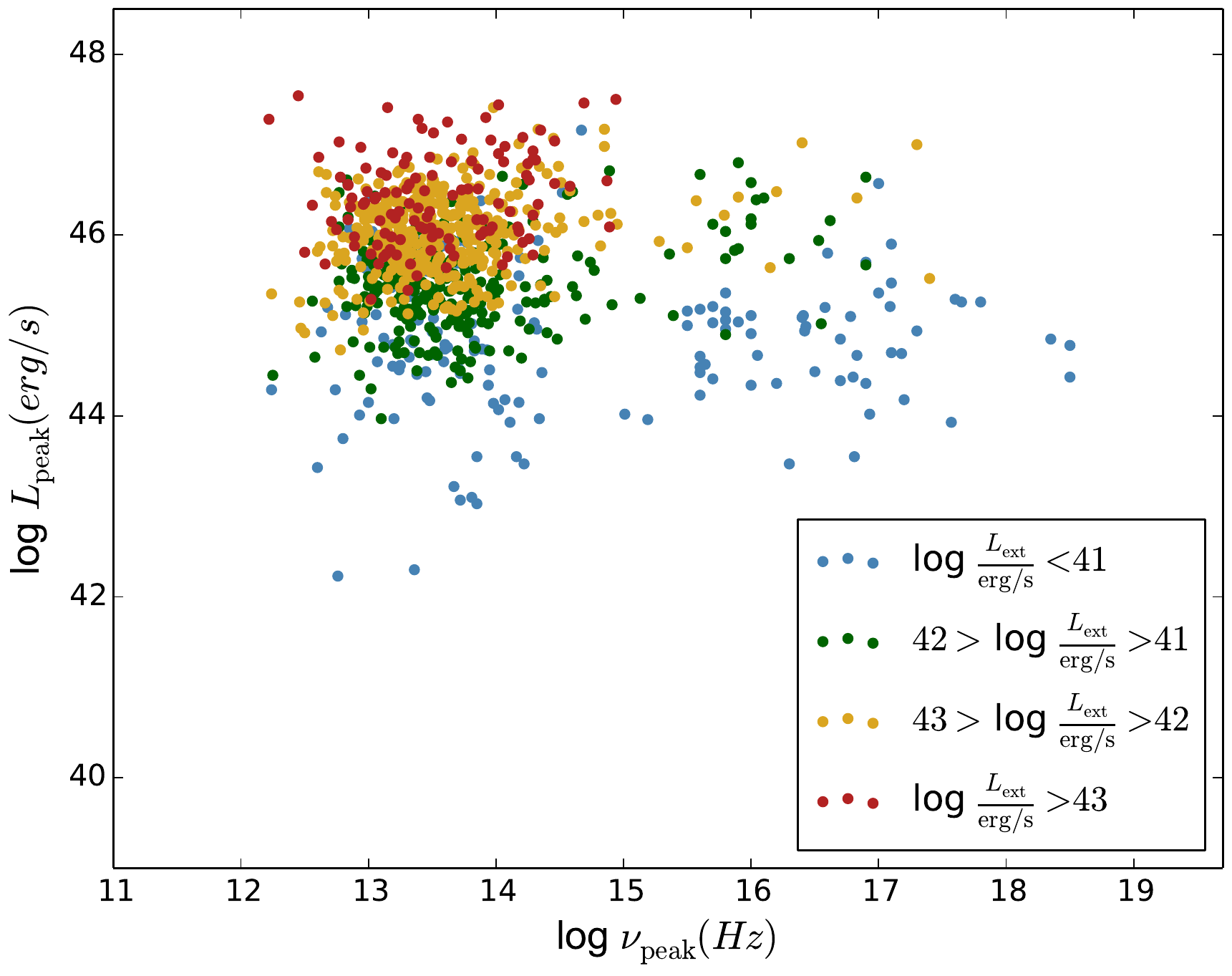}
    \caption{The UEX sample in the \vplp ~plane (excluding those with missing redshift information) with the color scale representing upper limit of $L_\mathrm{ext}$. We find that none of the high-$L_\mathrm{peak}$, high-$\nu_\mathrm{peak}$ sources have values in the two highest energy bin of Figure \ref{4panel} ($L_\mathrm{ext} > 10^{42}$ erg/s).}
    \label{fig:supp_uex_uls}
\end{figure*}

\begin{figure*}
    \centering
    \includegraphics[width=5in]{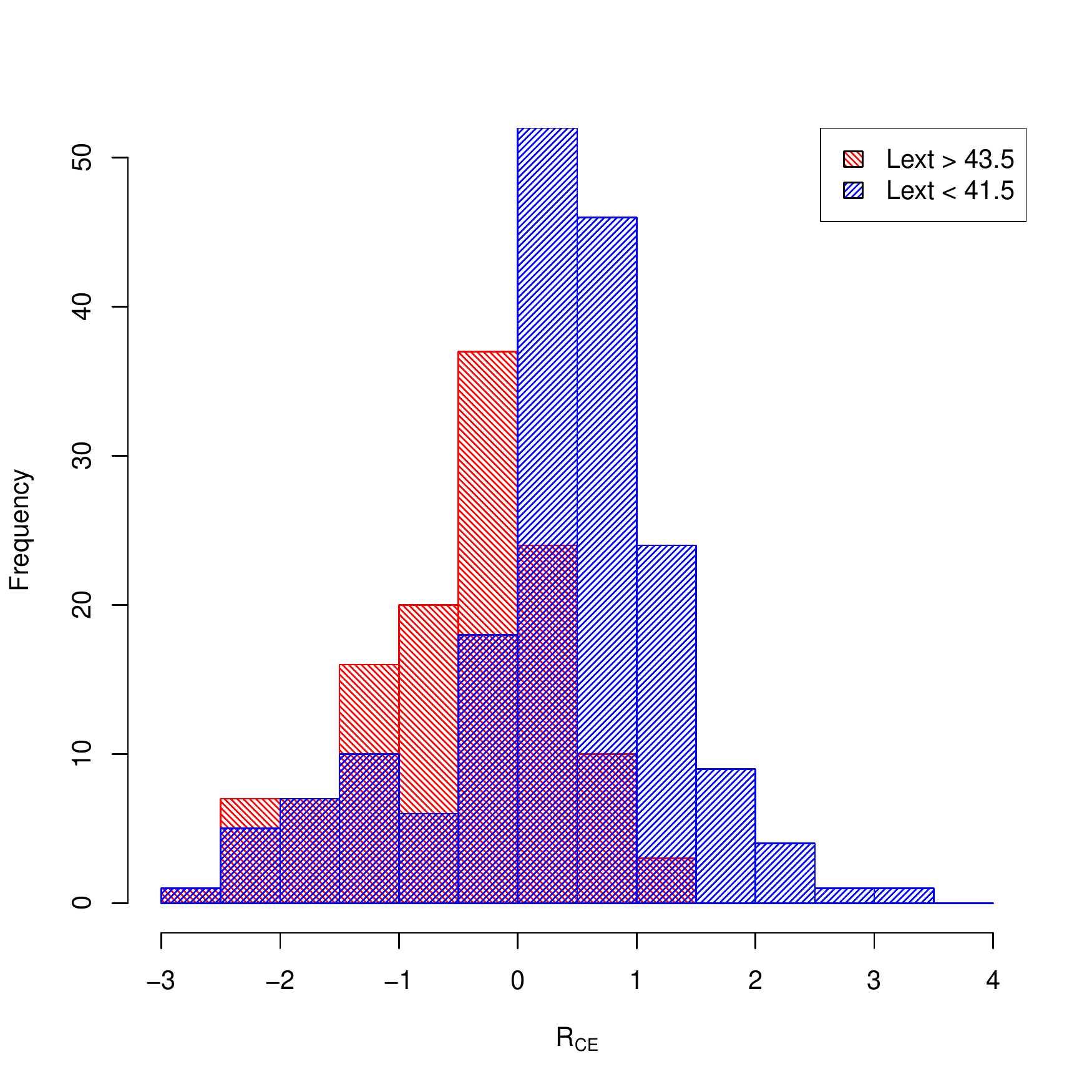}
    \caption{Comparison of the Radio Core Dominance Ranges for low and high $L_\mathrm{ext}$ sources in the TEX sample.}
    \label{supp:hist_rce}
\end{figure*}
\begin{figure*}
    \centering
    \includegraphics[width=5in]{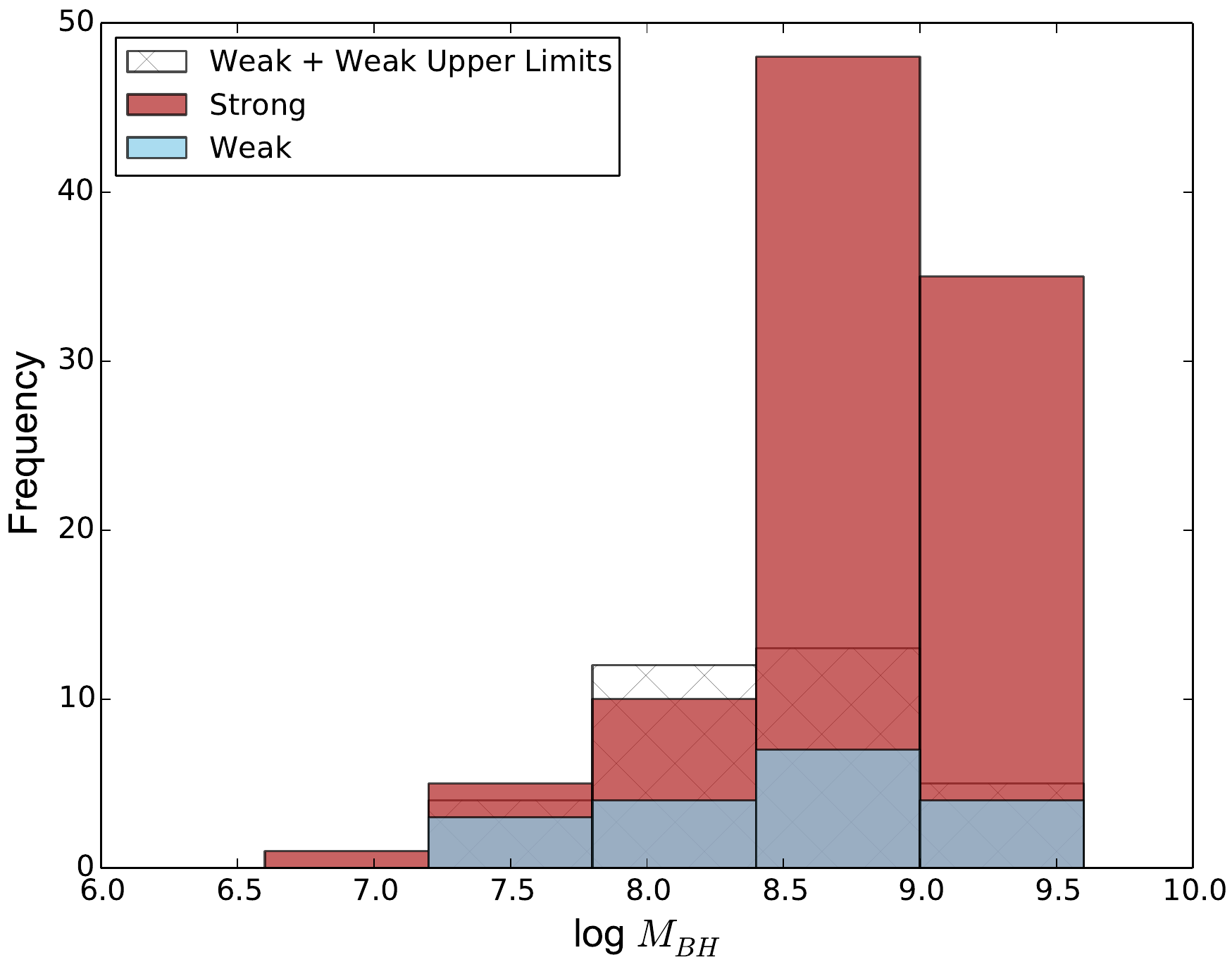}
    \caption{Histograms of the black hole masses for strong and weak jets (including the UEX sample for weak jets).}
    \label{supp:mbh_hist}
\end{figure*}

\begin{figure*}
    \centering
    \includegraphics[width=5in]{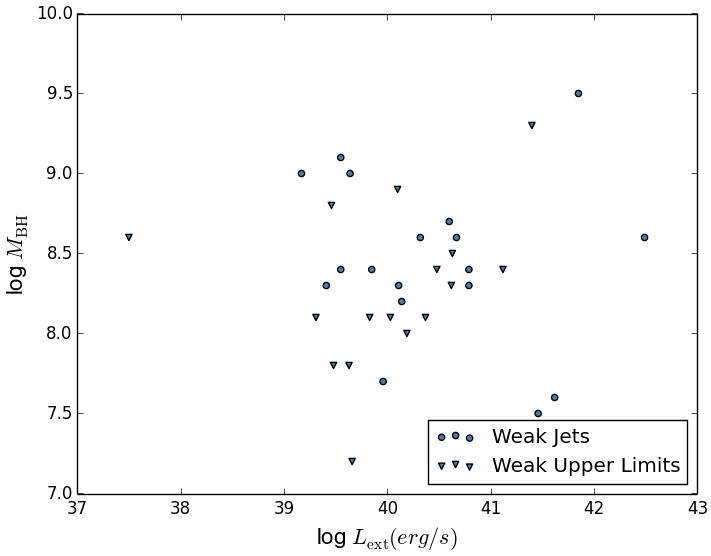}
    \caption{A plot of $M_\mathrm{BH}$ versus $L_\mathrm{ext}$ for weak jets.}
    \label{supp:lextmbh_weak}
\end{figure*}

\begin{figure*}
    \centering
    \includegraphics[width=5in]{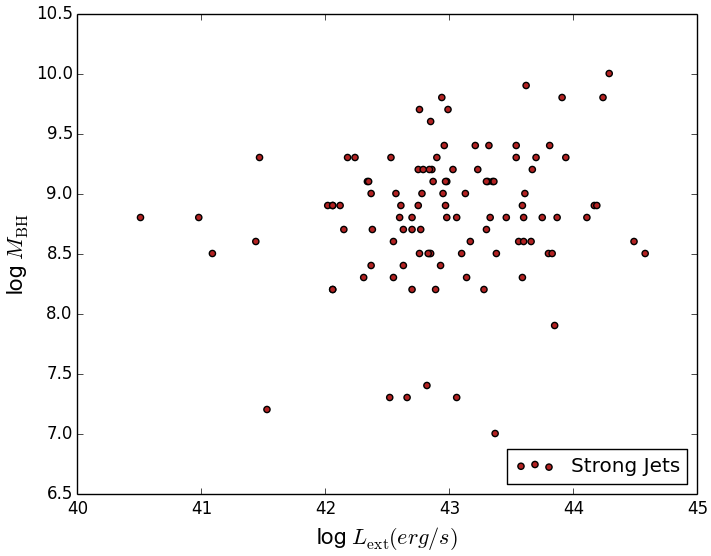}
    \caption{A plot of $M_\mathrm{BH}$ versus $L_\mathrm{ext}$ for strong jets.}
    \label{supp:lextmbh_strong}
\end{figure*}

%%%%%%%%%%%%%%%%%%%%%%%%%%%%%%%%%%%%%%%%%%%%%%%%%%

% Don't change these lines
\bsp	% typesetting comment
\label{lastpage}
\end{document}